# Reed-Solomon and Concatenated Codes with Applications in Space Communication

Polykarpos Thomadakis, Antonios Argyriou

University of Thessaly

## ABSTRACT

In this paper we provide a detailed description of Reed-Solomon (RS) codes, the most important algorithms for decoding them, and their use in concatenated coding systems for space applications. In the current literature there is scattered information regarding the bit-level implementation of such codes for either space systems or any other type of application. Consequently, we start with a general overview of the channel coding systems used in space communications and then we focus in the finest details. We first present a detailed description of the required algebra of RS codes with detailed examples. Next, the steps of the encoding and decoding algorithms are described with detail and again with additional examples. Next, we focus on a particularly important class of concatenated encoders/decoders namely the Consultative Committee for Space Data Systems (CCSDS) concatenated coding system that uses RS as the outer code, and a convolutional inner code. Finally, we perform a thorough performance evaluation of the presented codes under the AWGN channel model.

## 1. INTRODUCTION

The continuous search for knowledge of our universe and its origins is increasing as technology develops. This search for knowledge has led us to hundreds of space missions since the late 1950. Valuable information is collected through these missions, information that could not be collected from earth. To achieve this, however, several challenges have to be addressed by scientists and engineers.

During each deep space mission, reliable communication with the spacecraft, to send telecommands or software updates, track location and receive telemetry, images and scientific data is vital to the success of the mission. This need for reliable communication required the need for constant research and development in this area by modifying existing technologies or the invention of new ones. There is a number of factors that aggravate this effort and have to be fought in order to accomplish a successful space mission.

The first problem that is present in deep space communication is the enormous distances that the systems encounter. Earth communications cannot even approach these distances, so this is a new problem that has to be solved. Another important drawback is the very high latency constrained by the speed of light. The latency in such communications can even reach a period of days which would be prohibitive in common communications. In addition to that, the rates in these distances have to be low, much lower than those on Earth to maintain the message detectable and retrievable. Certainly, as the performance goals we set get higher the requirements to attain them get higher, as a result the development and research has to be constant. Other constraints met in space communications are the need for low power consumption, the size and weight, as the spacecraft may need to stay in space for a very long period of time and it has to be active and ready to receive messages the continuously. Appropriate hardware has to be designed with the least possible size and weight.

The space environment itself of course is not friendly for transmissions as errors are prevalent in the messages and so error detection and possibly correction is applied to receive the correct messages in either uplink or downlink. In order to achieve near error-free communication, a number of methods have been developed over the last few decades. The most common form of error control used in space missions is Forward Error-Correction (FEC). This method sends additional bits (parity), which can be used to check the consistency of the received data and then rebuild parts of the data stream if required.

Some of the most well-known codes that are put into use for this reason are convolutional codes, Reed-Solomon, the concatenation of those two, Turbo codes and most recently LDPC codes. In this paper we will focus on the theory and implementation of the concatenation of convolutional and Reed-Solomon codes. LDPC and Turbo codes were put in use more recently and seem to have a better performance but still several of the flying space missions in process use this concatenated system and it provides a very good performance. Furthermore, concatenation is an important channel coding technique that we describe thoroughly in this work.

The rest of this paper is organized as follows: In section 2 the required algebra used by Reed-Solomon is presented, while section 3 consists of the encoding and decoding algorithms with detailed examples. In section 4 the basic concept of convolutional codes is presented. In section 5 the implementation of the concatenated system is presented, and also its performance for different coding configurations. Finally, we conclude this paper in section 6.

## 2. GALOIS FIELD ALGEBRA

Galois field (GF) algebra is similar to conventional algebra except that GF algebra operates within a finite field. In GF algebra it is possible to take an element, sum with another element and obtain the resulting element only within a finite number of elements. There are some standard algebraic laws that govern GF; these laws will be first presented to understand GF arithmetic. GF is used in most block error correction codes one of which will be presented and used later.

### 2.1 Groups

Let G be a set of elements. A binary operation * on G is a rule that assigns to each pair of elements A and B a uniquely defined third element C=A*B in G. When such an operation is defined, G is closed under *. This operation * is called associative if, for any A , B and C in G

A*(B*C) = (A*B)*C.

*Definition 1:*

A set G (on which a binary operation is defined) is defined to be a group if the following conditions are satisfied:

1. The binary operation * is associative
2. G contains an identity element I such that, for any A in G, A*I = I*A = A.
3. For any element A in G, there exists an inverse element A' in G such that A*A' = A'*A = I.

A group G is commutative if its binary operation * also satisfies the following condition:

A*B = B*A, for all A and B in G

This is all we need to know about groups to perform GF arithmetic.

**2.2 Fields**

A field is a set of elements in which we can do addition, subtraction, multiplication, and division without leaving the set. Addition and multiplication must satisfy the commutative, associative, and distributive laws.

*Definition 2:*

A set F together with the two binary operations "+" and "-'" is a field if the following conditions are satisfied:

1.  F is a commutative group under addition "+". The identity element with respect to addition I is called the zero element or the additive identity I of F and is denoted by 0 (zero).

2.  F is a commutative group under multiplication "·'". The identity element with respect to multiplication I is called the unit (or unity) element or the multiplicative identity I of F and is denoted by 1 (one).

3.  Multiplication "'·" is distributive over addition "+"; that is, for any three elements A, B and C in F: A'(B+C) = (A ·B) + (A ·C).

So a field consists of at least 2 elements the additive identity element and the multiplication identity element. The number of elements in a field is called the order of the field. A field with a finite number of elements is called a finite field. For every element A in a field, A·0 = 0·A = 0. If A is non-zero and B is non-zero the A·B is non-zero.

**2.3 Binary Field GF (2)**

Consider the set of two integers, G= {0, 1}. Let us define a binary operation, denoted as modulo-2 addition "+", on G as follows:

**Table 2-2-1: Addition over GF(2)**

| + | 0 | 1 |
|---|---|---|
| **0** | 0 | 1 |
| **1** | 1 | 0 |

It can be proved that this is a group. It is closed and associative under "+". The additive element is 0, 0+1=1 and 0+0=0 .The additive inverse for 0 is 0 and for 1 it is 1, 0+0=0 and 1+1=0. So it can be also seen that 1+1=0 => 1=-1 and so 1+1=1-1=-1-1=0.As a result, addition and subtraction are equivalent in GF(2).

Consider the same set of two integers, F= {0, 1}. Let us define another binary operation, denoted as multiplication "·", on F as follows:

**Table 2-2-2: Multiplication over GF(2)**

| · | 0 | 1 |
|---|---|---|
| **0** | 0 | 0 |
| **1** | 0 | 1 |

It can be proven that this is a field under modulo-2 addition and multiplication. As previously shown F=G is commutative under addition 0 is the addition identity. Also the non-zero elements form a commutative group under multiplication. 1 is the multiplicative identity and 1 is the multiplicative inverse element since 1·1=1. In addition, multiplication is distributive over modulo-2 addition A· (B+C) = (A·B) + (A·C), thus F is a field. This modulo-2 field is the minimum field of finite number of elements. This field is usually called a binary or 2-ary field and it is denoted by GF(2). The binary field GF(2) plays a crucial role in error correction coding theory and is widely used in digital data transmission and storage systems.

## 2.4 Extension Fields GF($2^m$)

We are interested in prime finite fields called Galois fields GF(P). In the previous binary operation example the minimum number of possible elements was presented which comprised GF(2). Extension fields are GF($P^m$) where m=2,3,4,.. and P is prime. With the design of error correction coding based systems, we are interested in binary operations. Therefore, the focus is mainly on binary Galois fields GF(2) and the extended binary Galois fields GF($2^m$) from now on.

### 2.4.1 Primitive Polynomials p(x)

Polynomials over the binary field GF(2) are any polynomial with binary coefficients. These polynomials are produced by their factors e.g f(x) = f0 · f1· f2 ·...fk. A primitive polynomial p(x) can produce an extension field. It has to be an irreducible binary polynomial of degree m which divides $X^n$, where n=$P^m$-1 = $2^m$-1 and which does not divide $X^i$ for i<n. Any primitive polynomial p(X) can construct the $2^m$ unique elements including a 0 (zero or null) element and a 1 (one or unity) element. A degree m polynomial f(X) over GF($2^m$) is defined to be irreducible over GF($2^m$) if f(X) is not divisible by any polynomial over GF($2^m$) of degree greater than zero, but less than m. An irreducible 3rd degree (cubic) polynomial generates an 8 x 3-bit symbol field; an irreducible 4th degree polynomial generates 16 x 4-bit symbol field; an irreducible 8th degree polynomial generates 256 x 8-bit symbol field etc.

Every element of a GF ($2^m$ ) field is a lower order polynomial of the field generating polynomial. Each low order polynomial element of the field is of the form: P(x) =$b_{n-1} x^{n-1} + \cdots + b_2 x^2 + b_1 x + b_0 x^0$ where the coefficients $b_0$to $b_{n-1}$are binary or decimal values.

E.g. $p(x) = x^4 + x + 1$ is a primitive polynomial of degree m=4 and can generate a GF($2^4$)=GF(16).

The primitive polynomial is irreducible so it does not have a real integer root. We can set the primitive element α=2 to be the root so that $p(\alpha) = \alpha^4 + \alpha + 1 = 0 => \alpha^4 = \alpha + 1$ since it was earlier mentioned that 1+1=0 => 1=-1. The first elements of the field are 0 and α. Populate the rest of the field by multiplying the previous non-zero element by α and substituting $\alpha^4$ for $\alpha + 1$ until the field elements start to repeat. The null and unity elements of GF($2^m$) are equal to those of GF(2) which are 0 and 1 respectively.

Note that the positions of the bits in the 4-bit symbols match the positions of $\alpha^3, \alpha^2, \alpha^1$and $\alpha^0$ in the table2-3 below. We should also note that 1+1=0 => 2=0, so 2α=0.

**Table 2-3: GF(16) elements using $p(\alpha) = \alpha^4 + \alpha + 1$**

| Decimal | Binary | Element Polynomial | α | Derivation |
|---|---|---|---|---|
| 0 | 0000 | 0 | 0 | First element =0 |

| | | | | | | |
|---|---|---|---|---|---|---|
| 1 | 0001 | | | | 1 | $\alpha^0$ | Second element=1 |
| 2 | 0010 | | | $\alpha$ | | $\alpha^1$ | $\alpha \cdot 1 = \alpha$ |
| 4 | 0100 | | $\alpha^2$ | | | $\alpha^2$ | $\alpha \cdot \alpha = \alpha^2$ |
| 8 | 1000 | $\alpha^3$ | | | | $\alpha^3$ | $\alpha \cdot \alpha^2 = \alpha^3$ |
| 3 | 0011 | | | $\alpha$ | 1 | $\alpha^4$ | $\alpha^3 \cdot \alpha = \alpha^4 = \alpha + 1$ |
| 6 | 0110 | | $\alpha^2$ | $\alpha$ | | $\alpha^5$ | $\alpha \cdot \alpha^4 = \alpha \cdot (\alpha + 1) = \alpha^2 + \alpha$ |
| 12 | 1100 | $\alpha^3$ | $\alpha^2$ | | | $\alpha^6$ | $\alpha \cdot \alpha^5 = \alpha \cdot (\alpha^2 + \alpha) = \alpha^3 + \alpha^2$ |
| 11 | 1011 | $\alpha^3$ | | $\alpha$ | 1 | $\alpha^7$ | $\alpha \cdot \alpha^6 = \alpha \cdot (\alpha^3 + \alpha^2) = \alpha^4 + \alpha^3 = \alpha^3 + \alpha + 1$ |
| 5 | 0101 | | $\alpha^2$ | | 1 | $\alpha^8$ | $\alpha \cdot \alpha^7 = \alpha \cdot (\alpha^3 + \alpha + 1) = \alpha^4 + \alpha^2 + \alpha = \alpha^2 + \cancel{2\alpha} + 1 = \alpha^2 + 1$ |
| 10 | 1010 | $\alpha^3$ | | $\alpha$ | | $\alpha^9$ | $a \cdot a^8 = a \cdot (a^2 + 1) = a^3 + a$ |
| 7 | 0111 | | $\alpha^2$ | $\alpha$ | 1 | $\alpha^{10}$ | $a \cdot a^9 = a \cdot (a^3 + a) = a^4 + a^2 = \alpha^2 + \alpha + 1$ |
| 14 | 1110 | $\alpha^3$ | $\alpha^2$ | $\alpha$ | | $\alpha^{11}$ | $\alpha \cdot \alpha^{10} = \alpha \cdot (\alpha^2 + \alpha + 1) = \alpha^3 + \alpha^2 + \alpha$ |
| 15 | 1111 | $\alpha^3$ | $\alpha^2$ | $\alpha$ | 1 | $\alpha^{12}$ | $\alpha \cdot \alpha^{11} = \alpha \cdot (\alpha^3 + \alpha^2 + \alpha) = \alpha^4 + \alpha^3 + \alpha^2 = \alpha^3 + \alpha^2 + \alpha + 1$ |
| 13 | 1101 | $\alpha^3$ | $\alpha^2$ | | 1 | $\alpha^{13}$ | $\alpha \cdot \alpha^{12} = \alpha \cdot (\alpha^3 + \alpha^2 + \alpha + 1) = \alpha^4 + \alpha^3 + \alpha^2 + \alpha = \alpha^3 + \alpha^2 + \cancel{2\alpha} + 1 = \alpha^3 + \alpha^2 + 1$ |
| 9 | 1001 | $\alpha^3$ | | | 1 | $\alpha^{14}$ | $\alpha \cdot \alpha^{13} = \alpha \cdot (\alpha^3 + \alpha^2 + 1) = \alpha^4 + \alpha^3 + \alpha = \alpha^3 + \cancel{2\alpha} + 1 = \alpha^3 + 1$ |

Notice that the recursive process repeats itself once we create more than the $2^m$ unique field elements. Let's show this repetition through examples.

- $\alpha^{15} = \alpha \cdot \alpha^{14} = \alpha \cdot (\alpha^3 + 1) = \alpha^4 + \alpha = \cancel{2\alpha} + 1 = 1 = \alpha^0$
- $\alpha^{16} = \alpha \cdot \alpha^{15} = \alpha \cdot \alpha^0 = \alpha$

- $\alpha^{17} = \alpha \cdot \alpha^{16} = \alpha \cdot \alpha = \alpha^2$
- $etc\ldots.$

All these representations are equivalent, the power representation is used on multiplications and addition uses the vector representation. These two representations are the most commonly used.

The most common way of generating the field elements α is by using α=2=(0010)=x as demonstrated here. However, different primitive elements can also be used to generate the field besides α=2 but only this is needed for our purposes in this paper.

### 2.4.2 Addition and subtraction over GF($2^m$)

Addition and subtraction over the extended field GF($2^m$) are performed by using exclusive-or operation on the element's vector representations. For example:

$$a^4 = 0011$$
$$\underline{+\ a^8 = 0101}$$
$$a^4 XOR\ a^8 = 0110$$

$$a^1 = 0010$$
$$\underline{-\ a^2 = 0100}$$
$$a^1 XOR\ a^2 = 0110$$

In the table below the results of addition (or subtraction) for all combinations of elements is presented under GF($2^4$).

**Table 2-4: Addition/Subtraction over GF(16)**

| +` | 1 | $\alpha$ | $\alpha^2$ | $\alpha^3$ | $\alpha^4$ | $\alpha^5$ | $\alpha^6$ | $\alpha^7$ | $\alpha^8$ | $\alpha^9$ | $\alpha^{10}$ | $\alpha^{11}$ | $\alpha^{12}$ | $\alpha^{13}$ | $\alpha^{14}$ |
|---|---|---|---|---|---|---|---|---|---|---|---|---|---|---|---|
| **1** | 0 | $\alpha^4$ | $\alpha^8$ | $\alpha^{14}$ | $\alpha$ | $\alpha^5$ | $\alpha^{13}$ | $\alpha^9$ | $\alpha^2$ | $\alpha^7$ | $\alpha^5$ | $\alpha^{12}$ | $\alpha^{11}$ | $\alpha^6$ | $\alpha^3$ |
| **$\alpha$** | $\alpha^4$ | 0 | $\alpha^5$ | $\alpha^9$ | 1 | $\alpha^2$ | $\alpha^{11}$ | $\alpha^{14}$ | $\alpha^{10}$ | $\alpha^3$ | $\alpha^8$ | $\alpha^6$ | $\alpha^{13}$ | $\alpha^{12}$ | $\alpha^7$ |
| **$\alpha^2$** | $\alpha^8$ | $\alpha^5$ | 0 | $\alpha^6$ | $\alpha^{10}$ | $\alpha$ | $\alpha^3$ | $\alpha^{12}$ | 1 | $\alpha^{11}$ | $\alpha^4$ | $\alpha^9$ | $\alpha^7$ | $\alpha^{14}$ | $\alpha^{13}$ |
| **$\alpha^3$** | $\alpha^{14}$ | $\alpha^9$ | $\alpha^6$ | 0 | $\alpha^7$ | $\alpha^{11}$ | $\alpha^2$ | $\alpha^4$ | $\alpha^{13}$ | $\alpha$ | $\alpha^{12}$ | $\alpha^5$ | $\alpha^{10}$ | $\alpha^8$ | 1 |
| **$\alpha^4$** | $\alpha$ | 1 | $\alpha^{10}$ | $\alpha^7$ | 0 | $\alpha^8$ | $\alpha^{12}$ | $\alpha^3$ | $\alpha^5$ | $\alpha^{14}$ | $\alpha^2$ | $\alpha^{13}$ | $\alpha^6$ | $\alpha^{11}$ | $\alpha^9$ |
| **$\alpha^5$** | $\alpha^5$ | $\alpha^2$ | $\alpha$ | $\alpha^{11}$ | $\alpha^8$ | 0 | $\alpha^9$ | $\alpha^{12}$ | $\alpha^4$ | $\alpha^6$ | 1 | $\alpha^3$ | $\alpha^{14}$ | $\alpha^7$ | $\alpha^{12}$ |
| **$\alpha^6$** | $\alpha^{13}$ | $\alpha^{11}$ | $\alpha^3$ | $\alpha^2$ | $\alpha^{12}$ | $\alpha^9$ | 0 | $\alpha^{10}$ | $\alpha^{14}$ | $\alpha^5$ | $\alpha^7$ | $\alpha$ | $\alpha^4$ | 1 | $\alpha^8$ |

| $\alpha^7$ | $\alpha^9$ | $\alpha^{14}$ | $\alpha^{12}$ | $\alpha^4$ | $\alpha^3$ | $\alpha^{12}$ | $\alpha^{10}$ | 0 | $\alpha^{11}$ | 1 | $\alpha^6$ | $\alpha^8$ | $\alpha^2$ | $\alpha^5$ | $\alpha$ |
|---|---|---|---|---|---|---|---|---|---|---|---|---|---|---|---|
| $\alpha^8$ | $\alpha^2$ | $\alpha^{10}$ | 1 | $\alpha^{13}$ | $\alpha^5$ | $\alpha^4$ | $\alpha^{14}$ | $\alpha^{11}$ | 0 | $\alpha^{12}$ | $\alpha$ | $\alpha^7$ | $\alpha^9$ | $\alpha^3$ | $\alpha^6$ |
| $\alpha^9$ | $\alpha^7$ | $\alpha^3$ | $\alpha^{11}$ | $\alpha$ | $\alpha^{14}$ | $\alpha^6$ | $\alpha^5$ | 1 | $\alpha^{12}$ | 0 | $\alpha^{13}$ | $\alpha^2$ | $\alpha^8$ | $\alpha^{10}$ | $\alpha^4$ |
| $\alpha^{10}$ | $\alpha^5$ | $\alpha^8$ | $\alpha^4$ | $\alpha^{12}$ | $\alpha^2$ | 1 | $\alpha^7$ | $\alpha^6$ | $\alpha$ | $\alpha^{13}$ | 0 | $\alpha^{14}$ | $\alpha^3$ | $\alpha^9$ | $\alpha^{11}$ |
| $\alpha^{11}$ | $\alpha^{12}$ | $\alpha^6$ | $\alpha^9$ | $\alpha^5$ | $\alpha^{13}$ | $\alpha^3$ | $\alpha$ | $\alpha^8$ | $\alpha^7$ | $\alpha^2$ | $\alpha^{14}$ | 0 | 1 | $\alpha^4$ | $\alpha^{10}$ |
| $\alpha^{12}$ | $\alpha^{11}$ | $\alpha^{13}$ | $\alpha^7$ | $\alpha^{10}$ | $\alpha^6$ | $\alpha^{14}$ | $\alpha^4$ | $\alpha^2$ | $\alpha^9$ | $\alpha^8$ | $\alpha^3$ | 1 | 0 | $\alpha$ | $\alpha^5$ |
| $\alpha^{13}$ | $\alpha^6$ | $\alpha^{12}$ | $\alpha^{14}$ | $\alpha^8$ | $\alpha^{11}$ | $\alpha^7$ | 1 | $\alpha^5$ | $\alpha^3$ | $\alpha^{10}$ | $\alpha^9$ | $\alpha^4$ | $\alpha$ | 0 | $\alpha^2$ |
| $\alpha^{14}$ | $\alpha^3$ | $\alpha^7$ | $\alpha^{13}$ | 1 | $\alpha^9$ | $\alpha^{12}$ | $\alpha^8$ | $\alpha$ | $\alpha^6$ | $\alpha^4$ | $\alpha^{11}$ | $\alpha^{10}$ | $\alpha^5$ | $\alpha^2$ | 0 |

### 2.4.3 Multiplication and division over GF($2^m$)

Multiplication over GF($2^m$) is performed using the exponential representation by summarizing the symbol's exponents modulo $2^{m-1}$. For example:

$$a^5 \cdot a^2 = a^{5+2} = a^7$$

$$a^5 \cdot a^{14} = a^{19 \bmod 15} = a^{14}$$

In the table below the results of multiplication for all combinations of elements is presented under GF($2^4$).

**Table 2-5: Multiplication over GF(16)**

| $\cdot$ | 1 | $\alpha$ | $\alpha^2$ | $\alpha^3$ | $\alpha^4$ | $\alpha^5$ | $\alpha^6$ | $\alpha^7$ | $\alpha^8$ | $\alpha^9$ | $\alpha^{10}$ | $\alpha^{11}$ | $\alpha^{12}$ | $\alpha^{13}$ | $\alpha^{14}$ |
|---|---|---|---|---|---|---|---|---|---|---|---|---|---|---|---|
| **1** | 1 | $\alpha$ | $\alpha^2$ | $\alpha^3$ | $\alpha^4$ | $\alpha^5$ | $\alpha^6$ | $\alpha^7$ | $\alpha^8$ | $\alpha^9$ | $\alpha^{10}$ | $\alpha^{11}$ | $\alpha^{12}$ | $\alpha^{13}$ | $\alpha^{14}$ |
| $\boldsymbol{\alpha}$ | $\alpha$ | $\alpha^2$ | $\alpha^3$ | $\alpha^4$ | $\alpha^5$ | $\alpha^6$ | $\alpha^7$ | $\alpha^8$ | $\alpha^9$ | $\alpha^{10}$ | $\alpha^{11}$ | $\alpha^{12}$ | $\alpha^{13}$ | $\alpha^{14}$ | 1 |
| $\boldsymbol{\alpha^2}$ | $\alpha^2$ | $\alpha^3$ | $\alpha^4$ | $\alpha^5$ | $\alpha^6$ | $\alpha^7$ | $\alpha^8$ | $\alpha^9$ | $\alpha^{10}$ | $\alpha^{11}$ | $\alpha^{12}$ | $\alpha^{13}$ | $\alpha^{14}$ | 1 | $\alpha$ |
| $\boldsymbol{\alpha^3}$ | $\alpha^3$ | $\alpha^4$ | $\alpha^5$ | $\alpha^6$ | $\alpha^7$ | $\alpha^8$ | $\alpha^9$ | $\alpha^{10}$ | $\alpha^{11}$ | $\alpha^{12}$ | $\alpha^{13}$ | $\alpha^{14}$ | 1 | $\alpha$ | $\alpha^2$ |
| $\boldsymbol{\alpha^4}$ | $\alpha^4$ | $\alpha^5$ | $\alpha^6$ | $\alpha^7$ | $\alpha^8$ | $\alpha^9$ | $\alpha^{10}$ | $\alpha^{11}$ | $\alpha^{12}$ | $\alpha^{13}$ | $\alpha^{14}$ | 1 | $\alpha$ | $\alpha^2$ | $\alpha^3$ |

| | | | | | | | | | | | | | | |
|---|---|---|---|---|---|---|---|---|---|---|---|---|---|---|
| $\boldsymbol{\alpha^5}$ | $\alpha^5$ | $\alpha^6$ | $\alpha^7$ | $\alpha^8$ | $\alpha^9$ | $\alpha^{10}$ | $\alpha^{11}$ | $\alpha^{12}$ | $\alpha^{13}$ | $\alpha^{14}$ | $1$ | $\alpha$ | $\alpha^2$ | $\alpha^3$ | $\alpha^4$ |
| $\boldsymbol{\alpha^6}$ | $\alpha^6$ | $\alpha^7$ | $\alpha^8$ | $\alpha^9$ | $\alpha^{10}$ | $\alpha^{11}$ | $\alpha^{12}$ | $\alpha^{13}$ | $\alpha^{14}$ | $1$ | $\alpha$ | $\alpha^2$ | $\alpha^3$ | $\alpha^4$ | $\alpha^5$ |
| $\boldsymbol{\alpha^7}$ | $\alpha^7$ | $\alpha^8$ | $\alpha^9$ | $\alpha^{10}$ | $\alpha^{11}$ | $\alpha^{12}$ | $\alpha^{13}$ | $\alpha^{14}$ | $1$ | $\alpha$ | $\alpha^2$ | $\alpha^3$ | $\alpha^4$ | $\alpha^5$ | $\alpha^6$ |
| $\boldsymbol{\alpha^8}$ | $\alpha^8$ | $\alpha^9$ | $\alpha^{10}$ | $\alpha^{11}$ | $\alpha^{12}$ | $\alpha^{13}$ | $\alpha^{14}$ | $1$ | $\alpha$ | $\alpha^2$ | $\alpha^3$ | $\alpha^4$ | $\alpha^5$ | $\alpha^6$ | $\alpha^7$ |
| $\boldsymbol{\alpha^9}$ | $\alpha^9$ | $\alpha^{10}$ | $\alpha^{11}$ | $\alpha^{12}$ | $\alpha^{13}$ | $\alpha^{14}$ | $1$ | $\alpha$ | $\alpha^2$ | $\alpha^3$ | $\alpha^4$ | $\alpha^5$ | $\alpha^6$ | $\alpha^7$ | $\alpha^8$ |
| $\boldsymbol{\alpha^{10}}$ | $\alpha^{10}$ | $\alpha^{11}$ | $\alpha^{12}$ | $\alpha^{13}$ | $\alpha^{14}$ | $1$ | $\alpha$ | $\alpha^2$ | $\alpha^3$ | $\alpha^4$ | $\alpha^5$ | $\alpha^6$ | $\alpha^7$ | $\alpha^8$ | $\alpha^9$ |
| $\boldsymbol{\alpha^{11}}$ | $\alpha^{11}$ | $\alpha^{12}$ | $\alpha^{13}$ | $\alpha^{14}$ | $1$ | $\alpha$ | $\alpha^2$ | $\alpha^3$ | $\alpha^4$ | $\alpha^5$ | $\alpha^6$ | $\alpha^7$ | $\alpha^8$ | $\alpha^9$ | $\alpha^{10}$ |
| $\boldsymbol{\alpha^{12}}$ | $\alpha^{12}$ | $\alpha^{13}$ | $\alpha^{14}$ | $1$ | $\alpha$ | $\alpha^2$ | $\alpha^3$ | $\alpha^4$ | $\alpha^5$ | $\alpha^6$ | $\alpha^7$ | $\alpha^8$ | $\alpha^9$ | $\alpha^{10}$ | $\alpha^{11}$ |
| $\boldsymbol{\alpha^{13}}$ | $\alpha^{13}$ | $\alpha^{14}$ | $1$ | $\alpha$ | $\alpha^2$ | $\alpha^3$ | $\alpha^4$ | $\alpha^5$ | $\alpha^6$ | $\alpha^7$ | $\alpha^8$ | $\alpha^9$ | $\alpha^{10}$ | $\alpha^{11}$ | $\alpha^{12}$ |
| $\boldsymbol{\alpha^{14}}$ | $\alpha^{14}$ | $1$ | $\alpha$ | $\alpha^2$ | $\alpha^3$ | $\alpha^4$ | $\alpha^5$ | $\alpha^6$ | $\alpha^7$ | $\alpha^8$ | $\alpha^9$ | $\alpha^{10}$ | $\alpha^{11}$ | $\alpha^{12}$ | $\alpha^{13}$ |

Division over GF($2^m$) is performed also by using the exponential representation by subtracting the exponent of the second dividing symbol modulo $2^{m-1}$. Another way to perform division is first to find the inverse of the denominator and then perform multiplication. The inverse element is the one that multiplication with it will result in 1. For example:

$$\frac{a^5}{a^2} = a^{5-2} = a^3$$

$$\frac{a^5}{a^{14}} = \alpha^{-9 \bmod 15} = a^6$$

$$or$$

$$\frac{a^5}{a^{14}} = a^5 \cdot a^{-14} = a^5 \cdot a^1 = a^6$$

We can easily find the inverse from Table 2-5 above by locating the 1s . The element on this row is the inverse of the element in that column and vice-versa.

e.g.: $\alpha^{-1}$ -> $\alpha^{14}$

$\alpha^{-2}$ -> $\alpha^{13}$ etc…

### 2.4.4 Polynomial Arithmetic

The polynomials are really only representations of bit patterns with the values of the exponents dictating the location of the bits and the coefficients specifying the values at those locations. The polynomials allow us to visualise the arithmetic and to apply mathematical rules to the operations that are performed.

## 3. REED-SOLOMON CODES

### 3.1 Reed-Solomon background

Reed-Solomon (RS) codes are non-binary, BCH, cyclic, linear block error correction codes. Linear block codes are block architecture, optional systematic structure, and all code words are sums of code words. It has a block length of $n$ symbols and a message length of $k$ symbols. If the code is systematic, then it also has an unaltered data field of $k$ symbols independent of the associated parity-check field of $n$-$k$ symbols. Cyclic codes are codes where the linear block codes properties are valid and any cyclic rotation of a codeword is also a codeword.

A BCH code is a cyclic polynomial code over a finite field with a particularly chosen generator polynomial. They have the same characteristics as other cyclic codes, but with an additional characteristic; BCH codes can fairly easily be implemented into systems with any error correction capability of t symbols along with particular choices of the message length of $k$ symbols and the block length of $n$ symbols.

Reed-Solomon codes are very powerful burst error correcting codes, which means that they can correct many errors occurring one after the other in bits. This happens because they use decimal symbols instead of binary and the correction is made on whole symbols rather than bits.

### 3.1.1 Reed-Solomon Codewords

A binary based, t error correcting, primitive RS code has the following parameters:

Block length: $n = 2^{m-1}$ symbols

 Number of parity-checks: $n - k = 2t$ symbols

Minimum distance: $d_{min} = 2t + 1$ symbols

In the above $m$ is the number of bits per symbol $n$ is the codeword length and $k$ is the message length. RS codewords are denoted with the expression RS [n, k]. 2 x parity symbols are required to correct a single symbol error. Therefore, 2t = 32 would correct a 16 symbol errors.

Figure 3-1 illustrates a RS codeword of length n and correcting capability t, formed from k message symbols and 2t parity check symbols.

**Figure 3-1: A Reed-Solomon codeword**

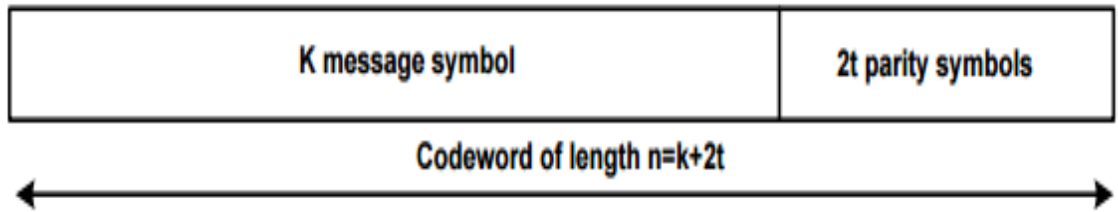

In decoding the RS codewords, essentially three events may happen.

a) The first event (correct decoding) happens if there are t or fewer RS symbol errors in a codeword. In this case the decoder successfully corrects the errors and outputs the correct information block.

b) The second event (detected error) happens if the number of RS symbol errors in a codeword is more than t, but the corrupted codeword is not close to any other codeword within the distance of t symbols. In this case the RS decoder fails to decode but can alert the user.

c) The third event (undetected error) happens if the number of RS symbol errors in a codeword is more than t, and the corrupted codeword is closer to some other codeword within the distance of t symbols. In this case the decoder is fooled, decodes incorrectly, and outputs a wrong information block. In other words, it claims the decoded block as a correct one and by doing this it may create up to t additional symbol errors (compared to the number of errors in the uncoded information block)

### 3.1.2 Reed-Solomon Polynomials

Two polynomials over GF($2^m$) are needed for the whole process. The field generator polynomial f(x) which produces the GF as we mentioned on chapter 2 and the generator polynomial g(x) of degree 2t-1 which is used to produce the parity check symbols.

### 3.2 Reed-Solomon encoder

The parity check information of 2t symbols is obtained from the message information

$$M(x) = M_{k-1}x^{k-1} + M_{k-2}x^{k-2} + \cdots + M_1x^1 + M_0 \qquad (3.1)$$

by dividing with the generator polynomial

$$g(x) = x^{2t} + g_{2t-1}x^{2t-1} + g_{2t-2}x^{2t-2} + \cdots + g_0 \qquad (3.2)$$

and taking the remainder (modulo function).

First we shift the message polynomial M(x) 2t symbols by multiplying with $x^{n-k} = x^{2t}$ then we divide with the generator polynomial g(x) and keep the remainder, the coefficients of the remainder are the parity check symbols. If CK(x) is the parity check:

$$CK(x) = x^{2t}M(x) \bmod g(x) = CK_{n-k-1}x^{n-k-1} + CK_{n-k-2}x^{n-k-2} + \cdots + CK_1x + CK_0$$
$$(3.3)$$

The codeword that is to be sent is produced by appending the parity check symbols to the transmitted message. This structure (taking the message and adding parity check symbols without changing the message symbols) is called systematic. The way to produce the codeword is to simply add the CK(x) polynomial of degree 2t-1 (length 2t) with the shifted version of the message polynomial which will have all zeroes in the last 2t symbols, thus no symbols will overlap from the two polynomials. The resulting codeword C will have a form like:

$$C(x) = M_{k-1}x^{n-1} + M_{k-2}x^{n-2} + \cdots + M_1x^{n-k+1} + M_0x^{n-k} + CK_{n-k-1}x^{n-k-1}$$
$$+ CK_{n-k-2}x^{n-k-2} + \cdots + CK_1x + CK_0$$
$$C(x) = C_{n-1}x^{n-1} + C_{n-2}x^{n-2} + \cdots + C_0 \qquad (3.4)$$

### 3.2.1 Generator Polynomial g(x)

The polynomial which produces the parity check information CK(x) to append to the message to be transmitted is the generator polynomial for a primitive RS code (of length $2^n - 1$ ) which is defined from the following equation:

$$g(x) = \prod_{i=FR}^{FR+2t-1} \left(x + a_g\right)^i$$
$$(3.5)$$

Where FR is the first root of the polynomial and $a_g$ is a primitive element of the field generator f(x). $a_g$ does not need to be the same element as the one used to produce the Galois field. It can be any primitive element of the field $a^k$, the most common implementations use FR=1 and $a_g = $ $\alpha$ as seen in bibliography. The choice of these parameters may result in different complexity in hardware design both in the encoder and the decoder.

The roots of a generator polynomial, g(x), must also be roots of the codeword generated by g(x), because a valid codeword is of the following form:

$$c(x) = q(x)\,g(x) \qquad (3.6)$$

where q(x) is a message-dependent polynomial. Therefore, an arbitrary codeword, when evaluated at any root of g(x), must yield zero, or in other words

$$g\big(a_g^i\big) = c\big(a_g^i\big) = 0 \qquad (3.7)$$

where i =FR,FR+1, FR+2, . . . , FR+2t-1.

A generator polynomial g(X) can also be constructed to be a self-reciprocating polynomial. Self-reciprocating polynomials have equivalent jth and i-jth coefficients, for example a reciprocating polynomial is

$$g(x) = x^6 + a^{10x^5} + a^{14}x^4 + a^4x^3 + a^{14}x^2 + a^{10}x + 1 \ (3.8)$$

The motive to use a self-reciprocating generator polynomial is that the encoder and decoder require less hardware.

### 3.2.2 An encoding example

Example 3.1:

Using as an example a RS(15,9) we will demonstrate the encoding process. Using the field generator

$$f(x) = x^4 + x + 1$$

from chapter 2 and the primitive element α to construct the field and code generator:

$$g(x) = \prod_{i=FR}^{FR+2t-1} \big(x + a_g\big)^i$$

For $a_g = \alpha$ and FR=1 we get:

$$g(x) = \prod_{i=1}^{2t} (x + a)^i = (x + a)(x + a^2)(x + a^3)(x + a^4)(x + a^5)(x + a^6) =$$

$$(x^2 + (a + a^2)x + a^3)(x^2 + (a^3 + a^4)x + a^7)(x^2 + (a^5 + a^6)x + a^{11})$$

$$= (x^2 + a^5x + a^3)(x^2 + a^7x + a^7)(x^2 + a^9x + a^{11})$$

$$= \cdots$$

$$g(x) = x^6 + a^{10}x^5 + a^{14}x^4 + a^4x^3 + a^6x^2 + a^9x + a^6$$

If m bits per symbol =4 and the message to be sent is

$$M(x) = 0x^8 + 0x^7 + 0x^6 + 0x^5 + 0x^4 + 0x^3 + 0x^2 + a^{11}x + 0 =$$

$$[0000000a^{11}0] = 0x0000000E0 \text{ in hexadecimal}$$

We must determine CK(x) first

$$CK(x) = x^6(a^{11}x)mod\,g(x)$$

$$CK(x) = x^6(a^{11}x)mod(x^6 + a^{10}x^5 + a^{14}x^4 + a^4x^3 + a^6x^2 + a^9x + a^6)$$

$$a^{11}\,x^7 \qquad\qquad \left| \dfrac{x^6 + a^{10}x^5 + a^{14}x^4 + a^4x^3 + a^6x^2 + a^9x + a^6}{a^{11}x + a^6} \right.$$

$$\underline{a^{11}x^7 + a^6x^6 + a^{10}x^5 + x^4 + a^2x^3 + a^5x^2 + a^2x} \qquad |$$

$$a^6x^6 + a^{10}x^5 + x^4 + a^2x^3 + a^5x^2 + a^2x \qquad |$$

$$\underline{a^6x^6 + ax^5 + a^5x^4 + a^{10}x^3 + a^{12}x^2 + x + a^{12}} \; |$$

$$= a^8x^5 + \; a^{10}x^4 + a^4x^3 + a^{14}x^2 + a^8x + a^{12}$$

So it is derived that the parity check polynomial CK is:

$$CK(x) = a^8x^5 + \; a^{10}x^4 + a^4x^3 + a^{14}x^2 + a^8x + a^{12}$$

Therefore, the codeword C(x) for our message M(x) is:

$$C(x) = x^6M(x) + CK(x)$$

$$C(x) = a^{11}x^7 + a^8x^5 + \; a^{10}x^4 + a^4x^3 + a^{14}x^2 + a^8x + a^{12}$$

Or

$$C = [0000000a^{11}0a^8a^{10}a^4a^{14}a^8a^{12}]$$

As we can see the result is a systematic codeword since the first nine symbols are the message symbols from M(x).

### 3.3 Reed-Solomon decoder

The decoding process is usually more complex and difficult to understand. In Reed-Solomon codes the procedure can be separated in 2 big steps:

1. Error detection part, in this part we calculate syndromes to detect whether there is an error in the receive codeword or not.

2. Error correction part, this part consists of four stages:

   - Find the error locator polynomial from the syndromes
   - Specify the error positions from the error locator polynomial
   - Calculate the error values from the syndromes and the error locator polynomial
   - Correct the errors found from the previous processes

The whole decoding process can be described from the figure below.

**Figure 3-2: Reed Solomon decoding process (adopted from [11])**

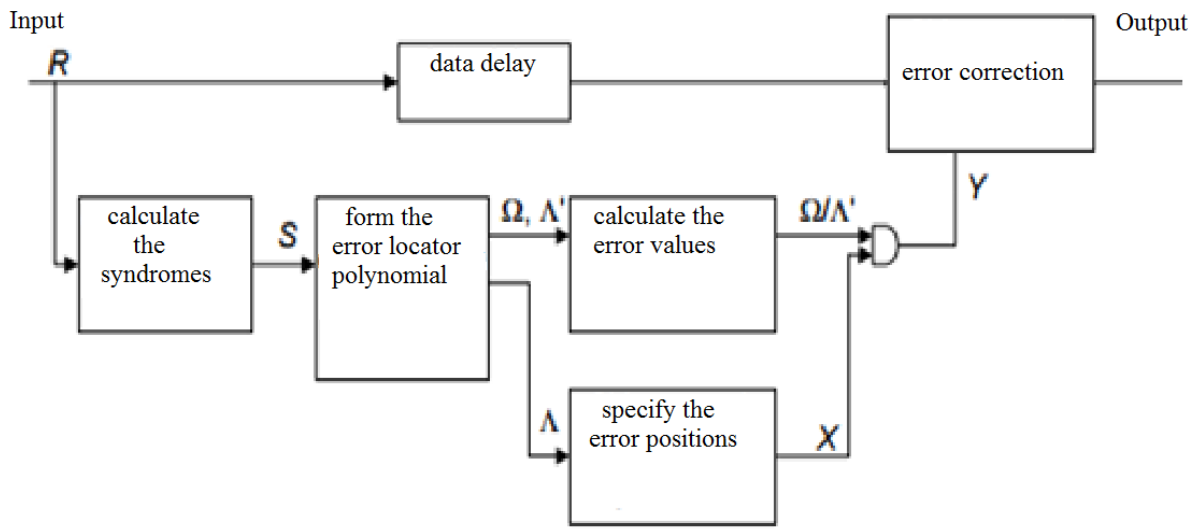

Each stage of the decoder will be explained in the following sections.

Before the explanation of each stage some definitions shall be made:

- Let the transmitted codeword polynomial be C(x) formed as follow:

$$C(x) = c_{n-1}x^{n-1} + \cdots + c_1 x + c_0 \,, where \ c_i \ is \ over \ GF(2^m) \qquad (3.8)$$

- Let the received codeword polynomial be R(x) formed as follow:

$$R(x) = r_{n-1}x^{n-1} + \cdots + r_1x + r_0 \text{ , where } r_i \text{ is over } GF(2^m) \quad (3.9)$$

- Let the error polynomial be E(x) added by the channel formed as :

$$E(x) = e_{n-1}x^{n-1} + \cdots + e_1 + e_0 \text{ , where } e_i \text{ is over } GF(2^m) \quad (3.10)$$

The received codeword R(x) is related with the others as shown below:

$$R(x) = C(x) + E(x) \quad (3.11)$$

In the decoding process we try to determine E(x) and then what we need to do is subtract the calculated E(x) from the received codeword R(x) to get the codeword C(x) and thus the message M(x) by removing the parity check symbols CK(x):

$$C(x) = R(x) - E(x) \quad (3.12)$$

$$M(x) = C(x) remove(CK(x)) \quad (3.13)$$

### 3.3.1 Syndrome calculation

Syndromes in the coding applications are some individual characteristics that characterize a particular error pattern. The syndrome polynomial S(x) is formed as:

$$S(x) = \sum_{i=FR}^{FR+2t-1} S_i x^{i-FR}$$

$$(3.14)$$

where FR is the first root of the generator polynomial g(x).

Each coefficient can be described as:

$$S_i = R(a_g^i),$$

$$i = FR, \dots, FR + 2t - 1 \quad (3.15)$$

If all coefficients are zero then there is no error, else if there is a non-zero coefficient it means there is an occurrence of error. For the previous RS(15,9) example 3.1:

Example 3.2:

$$C(x) = a^{11}x^7 + a^8x^5 + a^{10}x^4 + a^4x^3 + a^{14}x^2 + a^8x + a^{12}$$

by introducing some errors to this codeword we get the erroneous codeword R(x):

$$R(x) = x^8 + a^{11}x^7 + a^8x^5 + a^{10}x^4 + a^4x^3 + a^3x^2 + a^8x + a^{12}$$

Notice the errors on coefficients of $x^8$ and $x^2$, 0 has changed to 1 and $a^{14}$ to $a^3$ respectively. For $a_g = a^1$ and FR=1 we have:

$$S_1 = R(a^1) = a^8 + a^{11}(a^7) + a^8(a^5) + a^{10}(a^4) + a^4(a^3) + a^3(a^2) + a^8(a) + a^{12} = \cdots = 1$$

$$S_2 = R(a^2) = 1$$
$$S_3 = R(a^3) = a^5$$
$$S_4 = R(a^4) = 1$$
$$S_5 = R(a^5) = 0$$
$$S_6 = R(a^6) = a^{10}$$

The non-zero values indicate that there are errors in the codeword. The syndrome polynomial is formed:

$$S(x) = a^{10}x^5 + x^3 + a^5x^2 + x + 1$$

Another method to find the syndrome polynomial is to first find the remainder of R(x)/g(x), then evaluate the remainder polynomial for $a_g^i$ , i=FR,…,FR+2t-1 . The results are the coefficients of the syndrome polynomial in order from the lower x exponent to the highest.

### 3.3.2 Error locator polynomial

The next step, after the computing the syndrome polynomial is to calculate the error values and their respective locations. This stage involves the solving of the 2t syndrome polynomials, formed in the previous stage. These polynomials have T unknowns, where T is the number of unknown errors prior to decoding. If the unknown locations are $(i_1, i_2, …, i_T)$ the error polynomial can be expressed as,

$$E(x) = Y_1 x^{i_1} + Y_2 x^{i_2} + \cdots + Y_T x^{i_T} \qquad (3.16)$$

Where $Y_j$ is the magnitude of the jth error at location $i_j$. If $z_j$ is the field element associated with the error location $i_j$ , then the syndrome coefficients are given by,

$$S_i = \sum_{j=1}^{T} y_j z_j^i$$

$$(3.17)$$

where i=FR,FR+1,…,2t+FR-1.

The expansion of this sum (3.17) gives the following set of 2t equations in the T unknown locations $z_j$ and T unknown error magnitudes $y_j$.

$$S_1(x) = y_1 z_1 + y_2 z_2 + \cdots + y_T z_T$$

$$S_2(x) = y_1 z_1^2 + y_2 z_2^2 + \cdots + y_T z_T^2$$

$$\cdots$$

$$S_{2t}(x) = y_1 z_1^{2t} + y_2 z_2^{2t} + \cdots + y_T z_T^{2t}$$

$$(3.18)$$

The above set of equations must have at least one solution because of the way the syndromes are defined. This solution is unique. Thus the decoder's task is to find the unknowns given the syndromes. This is equivalent to the problem in solving a system of non-linear equations. Clearly, the direct solution of the system of nonlinear equations is too difficult for large values of T. For this we need to find the error locator polynomial. There are two different error locator polynomials which are related to each other. The degree of either of these polynomials determines the total number of error symbols T which is less than or equal to the error correction capability t.

The first one has the error locators $z_i \dots z_T$ as its roots, which means v factors of the form

$$(x + z_i), for\ i = 1 \dots T$$

$$\sigma(x) = (x + z_1)(x + z_2) \dots (x + z_T)$$

$$= x^T + \sigma_1 x^{T-1} + \cdots + \sigma_T$$

$$(3.19)$$

where T is the number of errors.

The alternative representation has the inverse of the error locators $z_i^{-1} \dots z_T^{-1}$ as its roots, so it has a form of

$$(1 + xz_i), for\ i = 1 \dots T$$

$$\Lambda(x) = (1 + xz_1)(1 + xz_2) \dots (1 + xz_T)$$

$$= 1 + \Lambda_1 x + \cdots + \Lambda_{T-1} x^{T-1} + \Lambda_T x^T$$

$$(3.20)$$

And the relation between them is:

$$\sigma(x) = x^T \Lambda(x^{-1})$$

$$(3.21)$$

So the coefficients $\Lambda_i$ and $\sigma_i$ are the same.

There are 2 commonly known methods to find the locator polynomial. Berklamp-Massey algorithm, the extended Euclidean algorithm for computing the GCD. We mainly focus on the extended Euclidean algorithm as it is the easiest to understand, but the other will be briefly described for completeness.

Our goal for each of the two algorithms is to solve the key equation

$$\Omega(x) \equiv \Lambda(x)S(x) \, mod \, x^{2t}$$

(3.22)

where $\Omega(x)$ is the error evaluator polynomial.

### 3.3.2.1 The Berlekamp-Massey algorithm

The Berlekamp-Massey algorithm relies on the fact that the matrix of equations is highly structured. This structure is used to obtain the vector σ by a method that is conceptually more complicated. If the vector $\Lambda$(x) is known, then the first row of the above matrix equation defines $S_{FR+T}$ in terms of $S_{FR}, S_{FR+1}, \ldots, S_{FR+T-1}$ .The second row defines $S_{FR+T+1}$ in terms of $S_{FR+1}, S_{FR+2}, \ldots, S_{FR+T}$ and so forth. This sequential process can be summarized by the recursive relation,

$$S_j = \sum_{i=1}^{T} \Lambda_i S_{j-1}$$

(3.23)

where j=FR+T, FR+T+1,…, FR+2T-1.

For fixed $\Lambda$, this is equivalent to the equation of an autoregressive filter. It can be implemented as a linear-feedback shift register that will consequently generate the known sequences of syndromes.A flowchart of the algorithm is presented in fig.3-3. So $\Lambda$(x) is derived from this process. If we need to calculate the $\Omega$(x) also, it is just as simple as multiplying $\Lambda$(x) with S(x).

**Figure 3-3-3: The Berlekamp-Massey algorithm**

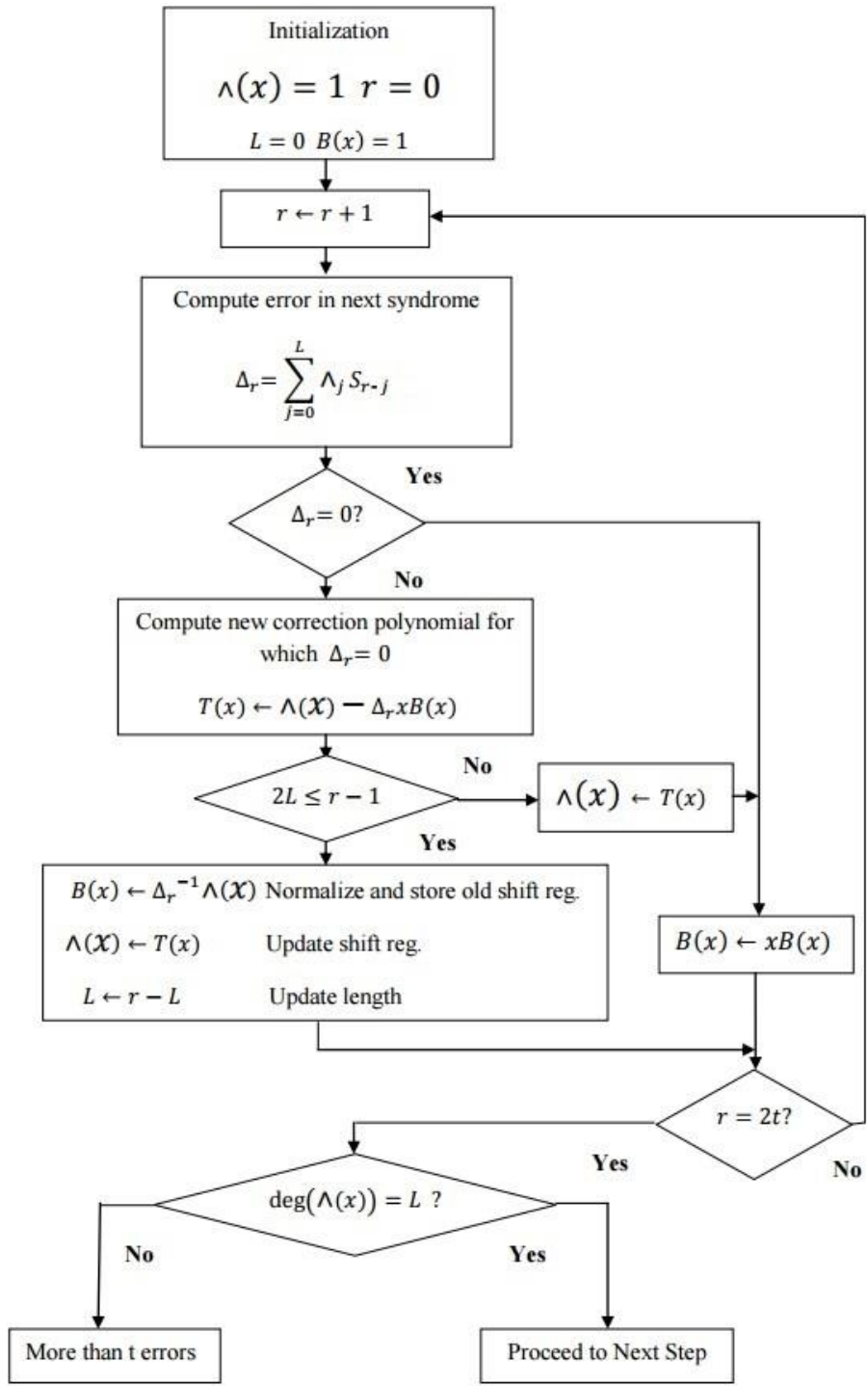

### 3.3.2.2 The extended Euclidean algorithm

The extended Euclidean algorithm (EEA) computes the greatest common divisor of two elements $a_1, a_2$ from a Euclidean domain E (e.g., a ring of polynomials over a field) and coefficients u, v ∈ E such that $a_1 u + a_2 v = \gcd(a_1, a_2)$. The algorithm proceeds by dividing $a_j$ by $a_{j+1}$ so that $a_j = a_{j+1} q_{j+1} + a_{j+2}$ with quotient $q_{j+1}$ and remainder $a_{j+2}$. Each step of the Euclidean algorithm works because the division implies that $gcd(a_j, a_{j+1}) = gcd(a_{j+1}, a_{j+2})$. For polynomials, the Euclidean algorithm terminates when $a_j = 0$. This always occurs because $deg(a_2) < deg(a_1)$ holds by assumption and $deg(a_{j+2}) < deg(a_1)$ holds by induction. The extended algorithm also computes $u_j$, $v_j$ recursively so that $a_j = u_j a_1 + v_j a_2$. Starting from $a_3 = a_1 - q_2 a_2$ ($i.e., u_3 = 1\ and\ v_3 = -q_2$), we have the recursion $a_{j+2} = a_j - q_{j+1} a_{j+1} = (u_j a_1 + v_j a_2) - q_{j+1}(u_{j+1} a_1 + v_{j+1} a_2)$. This gives the recursions

$$u_{j+2} = u_j - q_{j+1} u_{j+1} \rightarrow u_j = u_{j-2} - q_{j-1} u_{j-1} \quad (3.24)$$
$$and$$
$$v_{j+2} = v_j - q_{j+1} v_{j+1} \rightarrow v_j = v_{j-2} - q_{j-1} v_{j-1} \quad (3.25)$$

starting from $u_3 = 1$ and $v_3 = -q_2$.

The decoding of the RS codes is accomplished using a partial application of the EEA algorithm to compute $\gcd(x^{2t}, S(x))$. The extended part of the algorithm generates a sequence of relationships of the form:

$$u_j(x) x^{2t} + v_j(x) S(x) = a_j(x) \quad (3.26)$$

where the degree of $a_j(x)$ is decreasing with j. At the step jj where $deg\left(a_j(x)\right) < t$ for the first time the algorithm should stop. Viewing the above relationship as a congruence modulo $x^{2t}$ gives:

$$v_j(x) S(x) \equiv a_j(x)\ mod\ x^{2t} \quad (3.27)$$

So we see that $v_{jj}, a_{jj}$ satisfy the key equation for $v_{jj} = \Lambda(x)$ , $a_{jj} = \Omega(x)$. Another great advantage of the Euclidean algorithm is that we can receive both the error locator and the error evaluator polynomials from the same algorithm. A schematic representation of the whole algorithm can be seen below in fig.3-3. At the initialization process we set

$$\Lambda_{-1} = 0 \ ,$$
$$\Lambda_0 = 1$$
$$\Omega_{-1}(x) = x^{2t}$$
$$\Omega_0(x) = S(x)$$

Since we need to get the corresponding polynomials from the $\gcd\left(x^{2t}, S(x)\right)$ . $\Omega(x)$ is the remainder of the division on each step of the algorithm when its degree is lower than t the algorithm returns , giving us $\Lambda(x)$ and $\Omega(x)$ .

**Figure 3-4-4: The extended Euclidean algorithm**

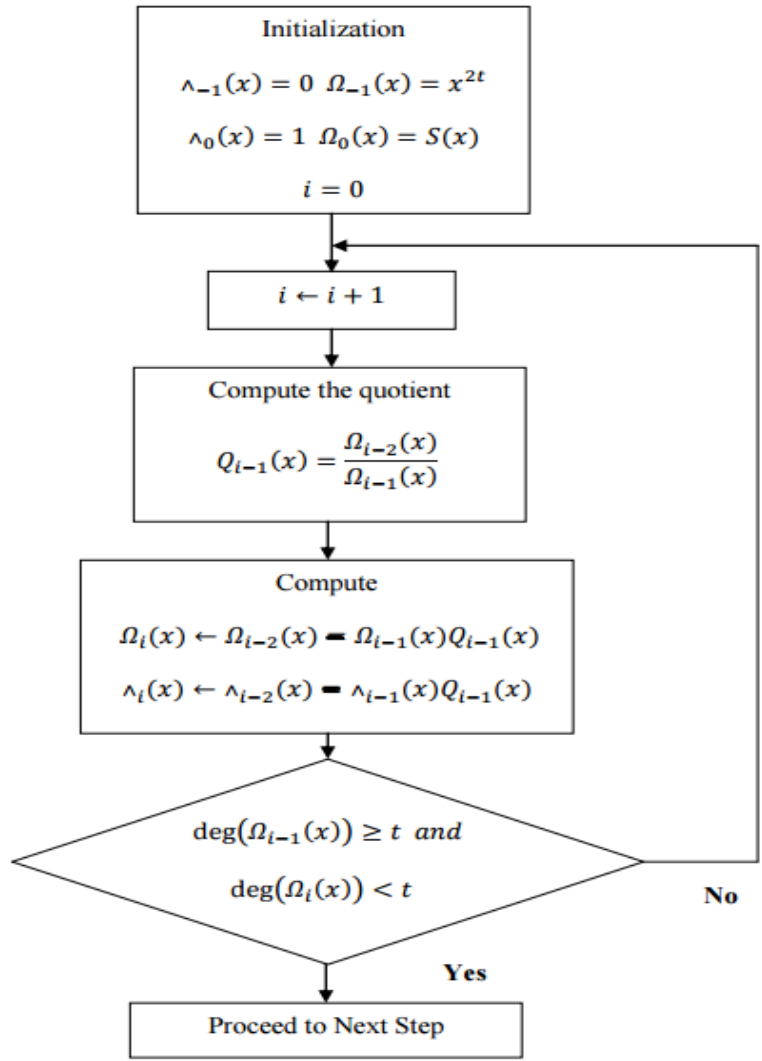

Example 3.3:

Let

$$S(x) = a^{10}x^5 + x^3 + a^5x^2 + x + 1$$

as the previous examples 3.1 and 3.2 , and t=3.

$$\Omega_{-1}(x) = x^{2t} = x^6$$
$$\Omega_0(x) = S(x)$$

August 13, 2016

Step 1:

Divide $x^{2t}$ by S(x):

$$x^{2t} = a^5 x S(x) + a^5 x^4 + a^{10} x^3 + a^5 x^2 + a^5 x$$

$$Q_0(x) = a^5 x$$

$$\Omega_1(x) = a^5 x^4 + a^{10} x^3 + a^5 x^2 + a^5 x$$

$$\Lambda_1(x) = Q_0(x) = a^5 x$$

The degree of $\Omega_1$ (=4) is bigger than t (=3) so the algorithm continues.

Step 2:

Divide $\Omega_0(x) = S(x)$ by $\Omega_1(x) = a^5 x^4 + a^{10} x^3 + a^5 x^2 + a^5 x$

$$S(x) = (a^5 X + a^{10}) \Omega_1(x) + 1$$

$$Q_1(x) = a^5 x + a^{10}$$

$$\Omega_2(x) = 1$$

$$\Lambda_2(x) = \Lambda_0(x) - \Lambda_1(x) Q_1(x) = 1 + (a^5 x)(a^5 x + a^{10}) = 1 + a^{10} x^2 + a^{15} x$$

$$\Lambda_2(x) = a^{10} x^2 + x + 1$$

The degree of $\Omega_2$ (=0) is smaller than t (=3) so the algorithm returns.

Therefore, the error locator polynomial is:

$$\Lambda(x) = a^{10} x^2 + x + 1$$

and the remainder

$$\Omega(x) = 1$$

is the error evaluator polynomial, which will be described in extent in the following sections.

### 3.3.3: Finding the error locations

Having calculated the error locator polynomial we are now able to locate the error positions. Our goal is to calculate the $z_i$ error locators using $\Lambda(x)$ and then the corresponding error locations $x_i$ in the erroneous received codeword. The algorithm used to determine these values is called Chien search.

### 3.3.3.1: The Chien search algorithm

The Chien search calculates the outputs for all the possible inputs; it is a very simple, brute force, search algorithm. The Chien search determines the roots of either the error-locator polynomial or of its reciprocal. The roots of the reciprocal of the error-locator polynomial σ(x) are the error-locator numbers $z_i$ and the <u>inverse</u> of the roots of the error-locator polynomial Λ(x) are the locator numbers $z_i$.

Using the Λ(x) polynomial it can be expressed as:

$$\Lambda(\text{x}) = \prod_{i=1}^{T} (x + z_i)$$

for $z_i = a^k$ for some k.                                                                                      (3.28)

All we need to do is to substitute into it the inverse of $a^n$ for each value of n in the R(x) codeword. The exponent location that returns a value of 0 is the locator number $z_i$. The reason for this is that S(x) contains the error information of the entire R(x) codeword and Λ(x) is derived using S(x). The non-erroneous coefficients do not contribute to the error component and return a non-zero value when substituted but the coefficient that is responsible for the error cancels out and returns 0. The error locations $x_i$ are defined from the error-locator numbers $z_i$ as $x_i = X^{\log_a z_i/G}$ where $G = \log_a a_g$. Using $a_g = a => G = 1$ but using $a_g = a^8 => G = 8$ We can also set the G to always be equal to 1 if instead of powers of $a$ we use powers of $a_g$ as inputs to the Chien search algorithm and the log base is changed to $a_g$.

For instance using the error locator polynomial computed in the previous step example 3.3,

Example 3.4:

$$\Lambda(x) = a^{10}x^2 + x + 1$$

for the erroneous received codeword

$$R(x) = x^8 + a^{11}x^7 + a^8x^5 + a^{10}x^4 + a^4x^3 + a^3x^2 + a^8x + a^{12}$$

produced by the correct codeword

$$C(x) = a^{11}x^7 + a^8 x^5 + a^{10}x^4 + a^4 x^3 + a^{14}x^2 + a^8 x + a^{12}$$

as previously, we get :

**Table 3-1: The outputs of Chien search**

| $\Lambda(a^0)$ | $a^{10} + 1 + 1$ | $a^{10}$ |
|---|---|---|
| $\Lambda(a^{-1})$ | $a^{10}(a^{-1})^2 + a^{-1} + 1$ | $a^{13}$ |
| $\Lambda(a^{-2})$ | $a^{10}(a^{-2})^2 + a^{-2} + 1$ | $0$ |
| $\Lambda(a^{-3})$ | $a^{10}(a^{-3})^2 + a^{-3} + 1$ | $a^{13}$ |
| $\Lambda(a^{-4})$ | $a^{10}(a^{-4})^2 + a^{-4} + 1$ | $a^7$ |
| $\Lambda(a^{-5})$ | $a^{10}(a^{-5})^2 + a^{-5} + 1$ | $a^{10}$ |
| $\Lambda(a^{-6})$ | $a^{10}(a^{-6})^2 + a^{-6} + 1$ | $a^5$ |
| $\Lambda(a^{-7})$ | $a^{10}(a^{-7})^2 + a^{-7} + 1$ | $a^9$ |
| $\Lambda(a^{-8})$ | $a^{10}(a^{-8})^2 + a^{-8} + 1$ | $0$ |
| $\Lambda(a^{-9})$ | $a^{10}(a^{-9})^2 + a^{-9} + 1$ | $a^5$ |
| $\Lambda(a^{-10})$ | $a^{10}(a^{-10})^2 + a^{-10} + 1$ | $a^0$ |
| $\Lambda(a^{-11})$ | $a^{10}(a^{-11})^2 + a^{-11} + 1$ | $a^9$ |
| $\Lambda(a^{-12})$ | $a^{10}(a^{-12})^2 + a^{-12} + 1$ | $a^7$ |
| $\Lambda(a^{-13})$ | $a^{10}(a^{-13})^2 + a^{-13} + 1$ | $a^6$ |
| $\Lambda(a^{-14})$ | $a^{10}(a^{-14})^2 + a^{-14} + 1$ | $a^6$ |

So the error locators are $z_1 = a^2$ and $z_2 = a^8$. The locations are $x_1 = X^{\log_a a^2/1} = X^2$ and $x_2 = X^{\log_a a^8/1} = X^8$. Which are indeed the location of the two errors as we can see comparing C(x) with R(x).

### 3.3.4: Calculating the error values

Two methods are available to calculate the error values, direct calculation and Forney's algorithm.

### 3.3.4.1: Direct calculation

We have already calculated the number of errors, their positions and their syndromes, so we can solve the following equation which was previously presented.

$$S_i = \sum_{j=1}^{T} y_j z_j^i$$

<div align="right">(3.29)</div>

All values are now known, except for y which is the error values. This would expand to a linear equations set easily solved now as we only need T=number of errors, of these equations to find y.

Using the RS(15,9) example again we only need $S_1 = 1$ and $S_2 = 1$ as T=2 and $z_1 = a^2, z_2 = a^8$ (calculated in examples 3.2 and 3.4 respectively) to find the error values, indeed :

<u>Example 3.5:</u>

$$S_1 = y_1 a^2 + y_2 a^8 = 1 \rightarrow y_1 a^2 + y_2 a^8 = 1$$
$$S_2 = y_1 (a^2)^2 + y_2 (a^8)^2 = 1 \rightarrow y_1 a^4 + y_2 a = 1$$

Solving this linear set results in $y_1 = 1, y_2 = 1$

### 3.3.4.2: Forney's algorithm

Methods of calculating the error values based on Forney's algorithm are more efficient than direct method solving the syndrome equations as described in the previous section. Forney's algorithm makes use of the calculated polynomials Λ(x) and Ω(x) and the error locator numbers $z_i$ to find these values. First of all it is needed to calculate the derivative of Λ(x) which a quite easy process. The derivative Λ'(x) is found by deleting the even powers of Λ(x) and dividing the result by x. The $y_i$ values are found by the following equation:

$$y_i = z_i^{1-FR} \left( \frac{\Omega(z_i^{-1})}{\Lambda'(z_i^{-1})} \right)$$

<div align="right">(3.30)</div>

When FR=1, the term $z_i^{1-FR}$ disappears, so the formula is often quoted in the literature as simply $\left(\frac{\Omega(z_i^{-1})}{\Lambda'(z_i^{-1})}\right)$, which gives wrong results for any other FR. Also in many books only the condition where FR=0 is presented so the formula is changed to $z_i\left(\frac{\Omega(z_i^{-1})}{\Lambda'(z_i^{-1})}\right)$ which is also giving the wrong impression about the general formula.

It should be noted that this equation only gives valid results for symbol locations that we know that they contain an error, if it is applied to other locations it will produce a mistake.

For the example we have used so far we have:

<u>Example 3.6:</u>

$$\Lambda'(x) = \frac{(a^{10}x^2 + x + 1)}{x} = 1$$
$$\Omega(x) = 1$$

We have found from Chien search that there are two errors so we need to calculate two y:

$$z_1 = a^2, z_2 = a^8$$
$$y_1 = a^{2^{1-1}}\left(\frac{\Omega(a^{-2})}{\Lambda'(a^{-2})}\right) = 1\left(\frac{1}{1}\right) = 1$$
$$y_2 = a^{8^{1-1}}\left(\frac{\Omega(a^{-8})}{\Lambda'(a^{-8})}\right) = 1\left(\frac{1}{1}\right) = 1$$

The error values are $y_1 = 1$, $y_2 = 1$, the same result as the direct calculation in example 3.5.

### 3.3.5 Decoding the codeword

So far the following steps have been described:

Step 1: Calculate the syndromes

Step 2: Find the error locator polynomial $\Lambda(x)$. Also find the error evaluator polynomial $\Omega(x)$.

Step 3: Perform the Chien Search to find the roots of $\Lambda(x)$.

Step 4: Find the magnitude of the error values using the Forney's Algorithm

One last step is left to complete the whole decoding process which is to use all that information that was extracted to actually decode the codeword. What needs to be done is to form the error polynomial E(x) and add (or subtract) it from the received codeword R(x). To achieve this we take the error values $y_i$ and place them in the location found by Chien search , $x_i = X^k$ the error polynomial will be of the form :

$$E(x) = \sum_{i=1}^{T} y_i x_i = y_1 x_1 + y_2 x_2 + \cdots + y_T x_T$$

The corrected codeword C'(x) is calculated by the following formula:

$$C'(x) = R(x) - E(x) = R(x) + E(x)$$

To obtain the message M(x) we simply strip the corrected codeword from the parity check symbols CK(x). If the errors occurred are less or equal to t we will have obtained the correct message.

To verify this with the example 3.1 used so far, we have:

Example 3.7:

$$R(x) = x^8 + a^{11}x^7 + a^8x^5 + a^{10}x^4 + a^4x^3 + a^3x^2 + a^8x + a^{12}$$
$$y_1 = 1 , y_2 = 1$$
$$x_1 = X^2$$
$$x_2 = X^8$$

$$E(x) = \sum_{i=1}^{T} y_i x_i = 1X^2 + 1X^8 = x^2 + x^8$$

Remember that the correct transmitted codeword was:

$$C(x) = a^{11}x^7 + a^8x^5 + a^{10}x^4 + a^4x^3 + a^{14}x^2 + a^8x + a^{12}$$

Adding E(x) with R(x) we get:

$$C'(x) = R(x) + E(x) =$$
$$= x^8 + a^{11}x^7 + a^8x^5 + a^{10}x^4 + a^4x^3 + a^3x^2 + a^8x + a^{12} + x^2 + x^8 =$$
$$= (1+1)x^8 + a^{11}x^7 + a^8x^5 + a^{10}x^4 + a^4x^3 + (a^3+1)x^2 + a^8x + a^{12} =$$
$$= a^{11}x^7 + a^8x^5 + a^{10}x^4 + a^4x^3 + (a^{14})x^2 + a^8x + a^{12}$$

Notice that C'(x) =C(x) the transmitted codeword. To receive the message itself we just strip the parity check bits from the corrected codeword. This is the last six symbols for our example since 2t=6.

So we get our decoded message

$$M(x) = a^{11}x$$

Which is indeed the message transmitted.

## 4. CONVOLUTIONAL CODES

Convolutional coding is another important coding technique. This method of coding is more used in communication channel with a memory. The implementation of convolutional codes is found in applications, which require good performance and low computational complexity. Unlike block codes, convolutional codes operate on code streams and the output does not depend only on the input bits, but also on previous bits. It converts any length of message to a single codeword. The name convolutional coding is given because the output bit stream of the encoder is convolution of input data bit and the transfer function of the encoder.

### 4.1 Convolutional encoder

The encoder of convolutional codes is formed by a linear finite-state shift register. A (n, k, L) convolutional code is represented by three parameters; n is the number of the encoder output bits, k is the number of bits as input to the encoder, L is called the constraint length and is related to the number of memory registers in the encoder. In general, a rate R=k/n, convolutional encoder input (information sequence) is a sequence of k binary bits,

$$u = u_0, u_1, u_2, \dots u_k \quad (4.1)$$

and the output (code sequence) is a sequence of n binary bits

$$c = c_0, c_1, c_2, \dots, c_n \quad (4.2)$$

The selection of which bits are to be added to produce the output bit is called the generator polynomial (g) for that output bit. The output sequence c, can be computed by convolving the input sequence u with the generator polynomial g and instead of normal addition modulo-2 addition is used:

$$c = u * g \quad (4.3)$$

A schematic representation of an encoder is presented in fig.5-1 below.

**Figure 4-1: A (2, 1, 4) convolutional code (adopted from [7])**

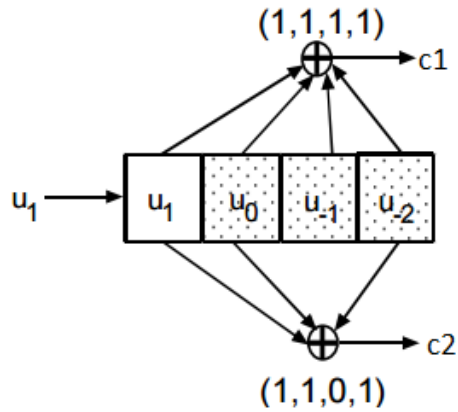

Here for each input bit 3 output bits are produced. The constraint length is L=4 and there are 2 generator polynomials $g_1 = (1\ 1\ 1\ 1)$ , $g_2 = (1\ 1\ 0\ 1)$ . So the outputs will be:

$$c_1 = mod2(u_1 + u_0 + u_{-1} + u_{-2})$$
$$c_2 = mod2(u_1 + u_0 + u_{-2})$$

So we see that g forms the output data.

## 4.2 States of a code

A convolutional coder has a finite number m of memory and consequently a finite number $2^{L-1}$ of memory states. These states play an important role on the output of the encoder as for each time the result of the encoder output depends on the current state. An example of the effect of the states can be seen in table 5-1 below for the coder represented in fig.5-1.

**Table 4-1: Look up table for the encoder of fig.4-1**

| Input | Current State | | | Output | | Next State | | |
|---|---|---|---|---|---|---|---|---|
| u | S1 | S2 | S3 | c1 | c2 | S1 | S2 | S3 |

| 0 | 0 | 0 | 0 | 0 | 0 | 0 | 0 | 0 |
|---|---|---|---|---|---|---|---|---|
| 1 | 0 | 0 | 0 | 1 | 1 | 1 | 0 | 0 |
| 0 | 0 | 0 | 1 | 1 | 1 | 0 | 0 | 0 |
| 1 | 0 | 0 | 1 | 0 | 0 | 1 | 0 | 0 |
| 0 | 0 | 1 | 0 | 1 | 0 | 0 | 0 | 1 |
| 1 | 0 | 1 | 0 | 0 | 1 | 1 | 0 | 1 |
| 0 | 0 | 1 | 1 | 0 | 1 | 0 | 0 | 1 |
| 1 | 0 | 1 | 1 | 1 | 0 | 1 | 0 | 1 |
| 0 | 1 | 0 | 0 | 1 | 1 | 0 | 1 | 0 |
| 1 | 1 | 0 | 0 | 0 | 0 | 1 | 1 | 0 |
| 0 | 1 | 0 | 1 | 0 | 0 | 0 | 1 | 0 |
| 1 | 1 | 0 | 1 | 1 | 1 | 1 | 1 | 0 |
| 0 | 1 | 1 | 0 | 0 | 1 | 0 | 1 | 1 |
| 1 | 1 | 1 | 0 | 1 | 0 | 1 | 1 | 1 |
| 0 | 1 | 1 | 1 | 1 | 0 | 0 | 1 | 1 |
| 1 | 1 | 1 | 1 | 0 | 1 | 1 | 1 | 1 |

As it is hard to retrieve the information easily from this table two ways of graphically representing them have been invented: State diagram and trellis diagram.

### 4.2.1 State diagram

A state diagram is nothing more than a finite state machine with the states of the encoder as its states , the input bits as its transition options and the output bits are represented on the transition arrows as shown below for the same convolutional coder as before.

**Figure 4-2: State diagram of the (2,1,4) code of fig.4-1 (adopted from [7])**

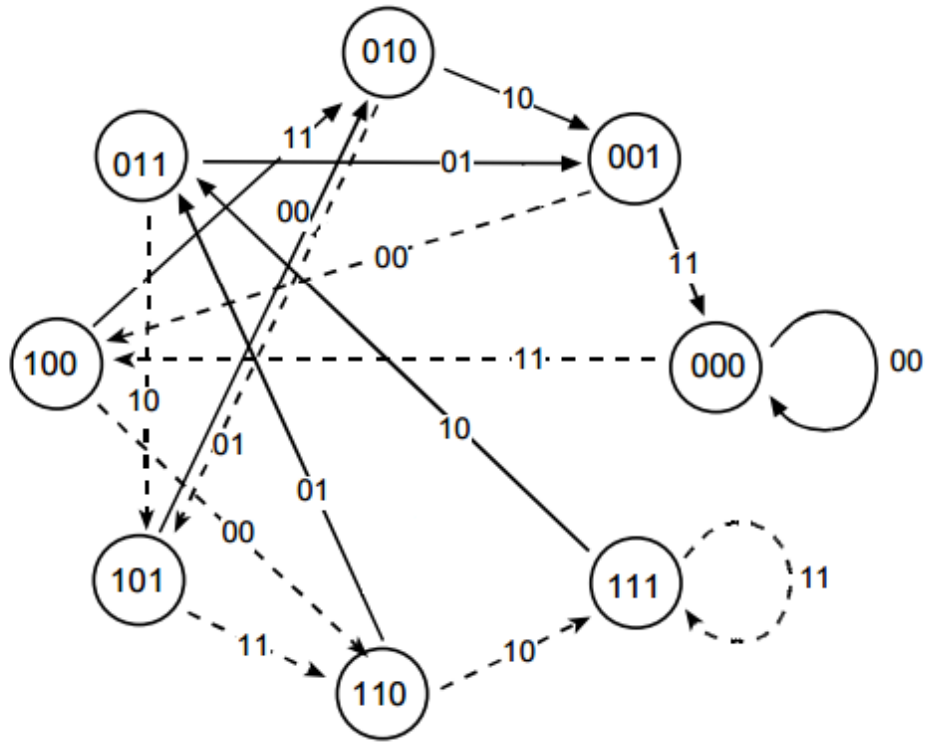

We obtain the same information here as the look up table at table 4-1. The solid lines indicate input bit 0 and the dashed ones input bit 1.

### 4.2.2 Trellis diagram

A trellis diagram is an extended representation of state diagram. For each instant of time it shows all the possible states. A unique path through the trellis represents the input bits and output bits. A trellis diagram consists of a node and the branches representing the state of the encoder and the transition of state. The initial node of the trellis diagram is the starting node. A combination of consecutive branches that connects the initial node to another node in the trellis is called a path and the number of branches comprising a path is called the length of the path. The trellis diagram is drawn by lining up all the possible states $2^{L-1}$ in the vertical axis and then connect each state to the next by the allowable codeword for that state. There are only two choices possible at each state, determined by the arrival of either bit 0 or bit 1. It is always assumed that the encoder is cleared and the initial state of the encoder is all zero. The arrows going upwards represent a 0 bit and those going downwards represent a 1 bit. The output is represented on the arrows as shown in fig.4-3.

**Figure 4-3: Trellis diagram of the (2,1,4) code (adopted from [7])**

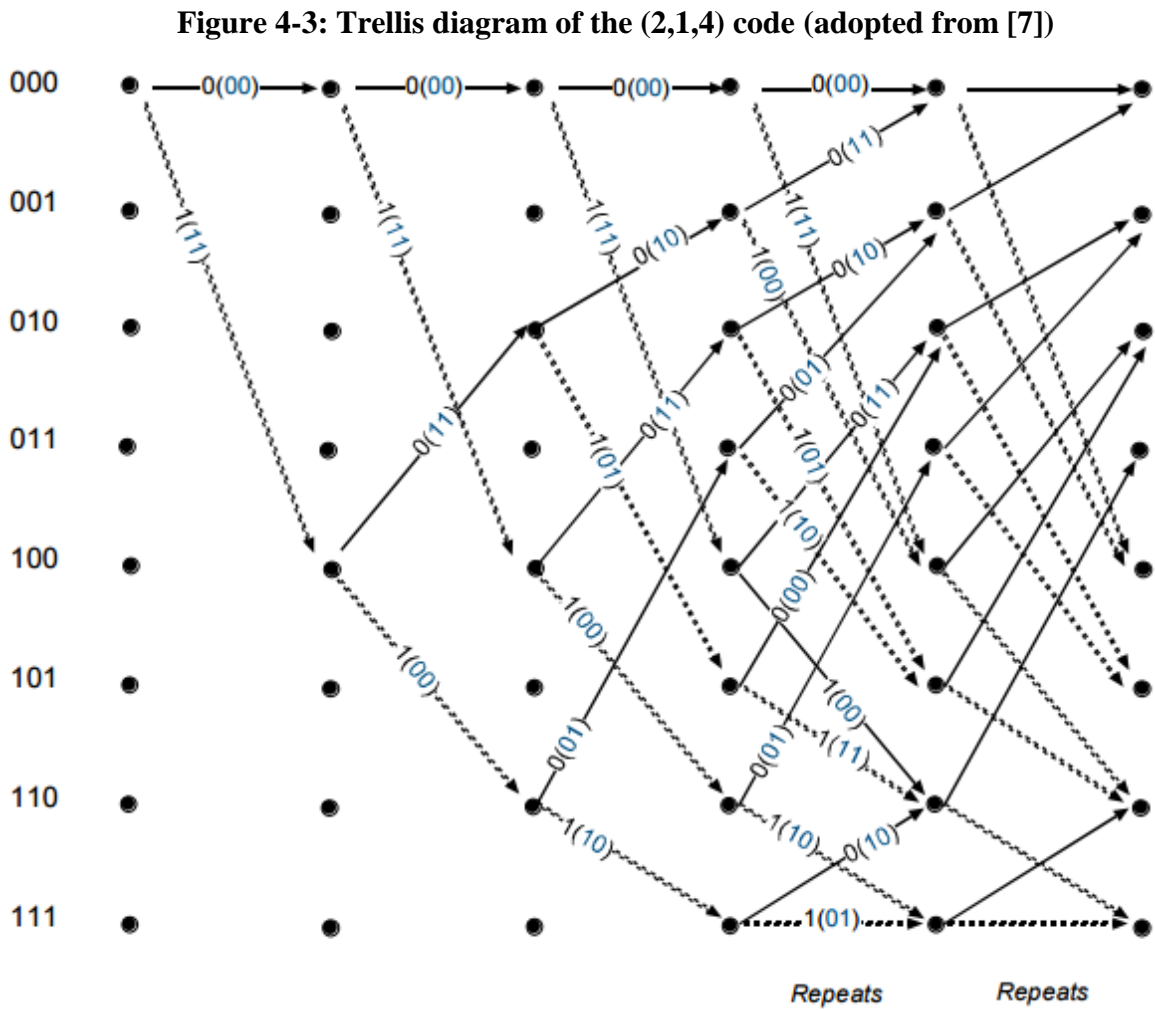

To encode a message we start at the 000 state and follow the arrows according to the message we have, the result sequence is the appending of the symbols on the corresponding arrows one after the other.

## 4.3 Convolutional decoder

The basic idea of decoding is that when we receive a codeword it might have errors or not. From the encoding process we know all the possible codewords for a specific number of bits which depend on standard bit inputs. What we try to do is to map erroneous codewords to the correct ones based on some criteria as how close it is to one of the correct codewords.

### 4.3.1 The Viterbi decoder

The Viterbi decoder examines an entire received sequence of given length. The decoder computes a metric for each  path of the trellis and makes a decision. All paths are computed until two paths converge on one node, in which case we keep the one with the higher metric (ties are broken arbitrarily).  We repeat this process for every stage, and at the end we add tail bits in order to force a return in the all-zero state. The surviving path is the decoder's decision.

The metric that can be used is usually hamming metric or Euclidean distance. Each of these metrics defines a different decoding decision making hard or soft respectively. Hamming metric denotes how many bits are the same between a codeword and the input the higher the better.

For example the (2,1,4) decoding process is presented in fig.4-4,4-5 for input of (01 11 01 11 01 01 11).

**Figure 4-4: The first 4 steps of Viterbi decoding of the (2,1,4) example (adopted from [7])**

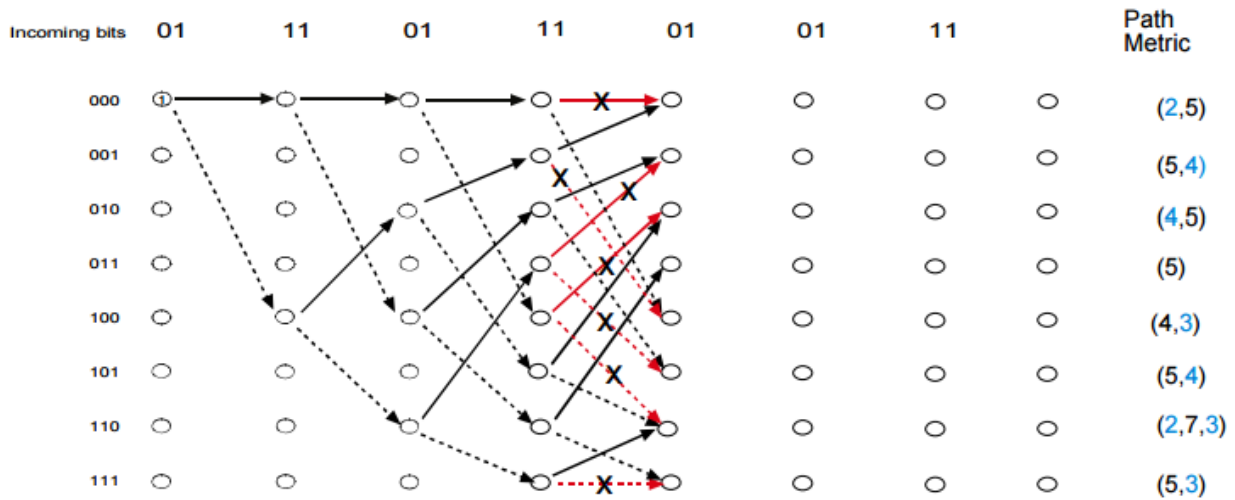

For the first 3 steps we just follow the arrows and accumulate the metrics for each path. At step 4 however, we need to discard some paths as they converge. The red lines crossed out show that these paths are discarded as their metric is lower. We only continue the process for the surviving paths. The result will be the following diagram in fig4-5.

**Figure 4-5: The final step of the decoding (adopted from [7])**

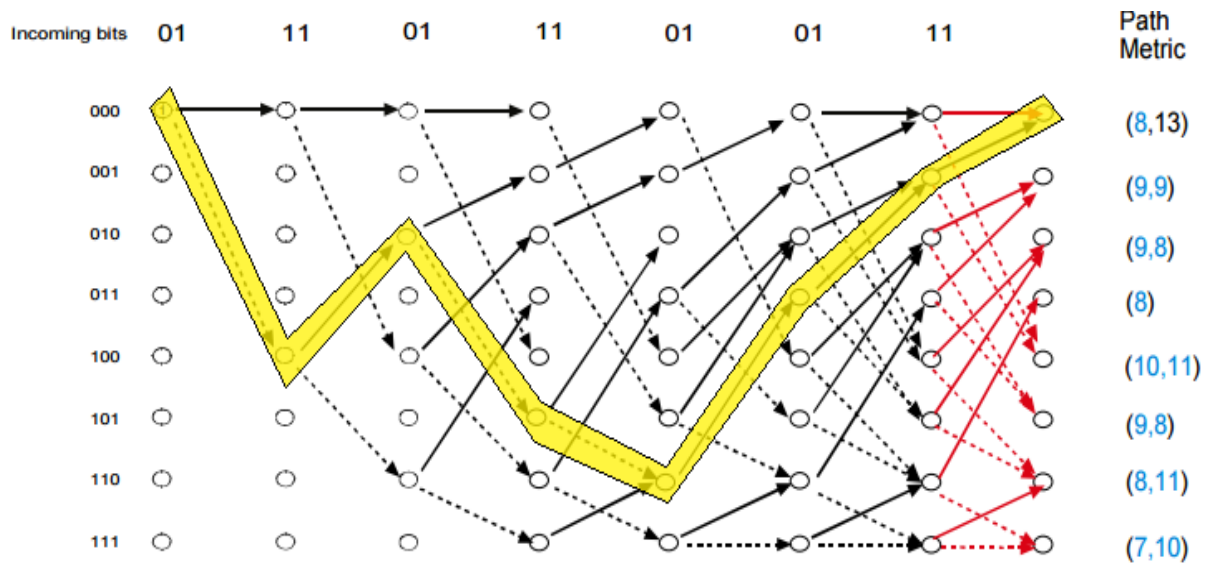

So the decoded bit sequence is 1011000. Note that the 4 zeroes in the end are there to return the decoder to all zero state.

In essence, we are performing Maximum Likelihood Sequence Estimation (MLSE), locating in the trellis the sequence most likely transmitted. We can also use the Viterbi algorithm with soft decoding. In this case we use the Euclidean distance instead of the Hamming distance. The Euclidean distance between two vectors r and s is:

$$d = \sum_i |r_i - s_i|^2$$

In this expression, r i are the unquantized values from the received sequence fed into the Viterbi decoder, and s i are the fixed values corresponding to a given sequence in the trellis. We also choose based on the lowest metric rather than the highest.

### 4.3.2 Truncation

In practice, we do not wait until the end of the source sequence to make a decision about which sequence was transmitted. The algorithm makes a decision about bits that are "sufficiently" in the past. In particular, at every stage we make a decision about which bits were transmitted before $L_{Tr}$ stages, before we move to the next stage. The parameter $L_{Tr}$ is called the truncation depth. In this way, we only retain in the memory of the decoder data about a sliding window, as shown below. A good choice has turned out to be $L_{Tr} \approx 6L$ (six constraint lengths).

## 5. SIMULATION AND PERFORMANCE RESULTS

### 5.1 Simulation approach

After successfully implementing the proposed decoder parameters in MATLAB simulations were performed to analyze their effectiveness. The channel used is an AWGN channel, the modulation is Binary phase-shift keying (BPSK) and 3 coding methods are tested. Convolutional codes, Reed Solomon codes and their concatenation. After that an attempt is made to increase even more the Bit error rate with usage of interleaving.

### 5.2 Simulation of the Convolutional code

The convolutional code can be described by the following figure fig.5-1.

**Figure 5-1: The convolutional encoder (adopted from [8])**

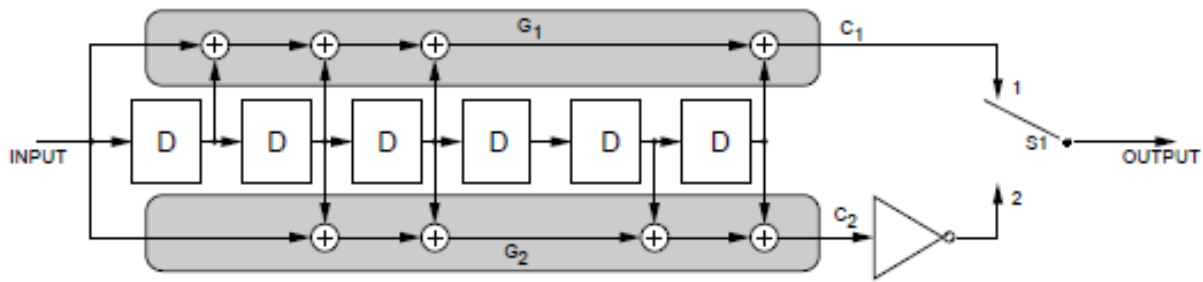

The characteristics of this coder are:

- Code rate:      ½ bit per symbol

- Constraint Length : 7 bits

- Generator vectors : G1=(1111001) , G2=(1011011)

A (7,1/2) convolutional code selected for space applications in the 1970s was a standout performer for its time. Exhaustive search over all convolutional codes with r=1/2 and K≤7 found that only this code (not counting a few symmetric equivalents) was able to achieve a free distance $d_{free}=10$

So it is a (2,1,7) encoder with an inventor in the second output bit.

The output sequence will be $c_1(1), \overline{c_2(2)}$

The decoding process uses soft maximum likelihood decision Viterbi decoder. With Viterbi decoding, it is possible to greatly reduce the effort required for maximum likelihood decoding by taking advantage of the special structure of the code trellis. Normally the decoder should operate on the entire received sequence, however, this is not achievable since the long latency and excessive memory storage required is too large. So a truncation length of 60 bits looks to be perfect for our purpose.

The performance of this code is presented in fig.5-2 compared to the uncoded BPSK version for the same channel and the same number of bits transmitted.

**Figure 5-2: Performance of the convolutional code**

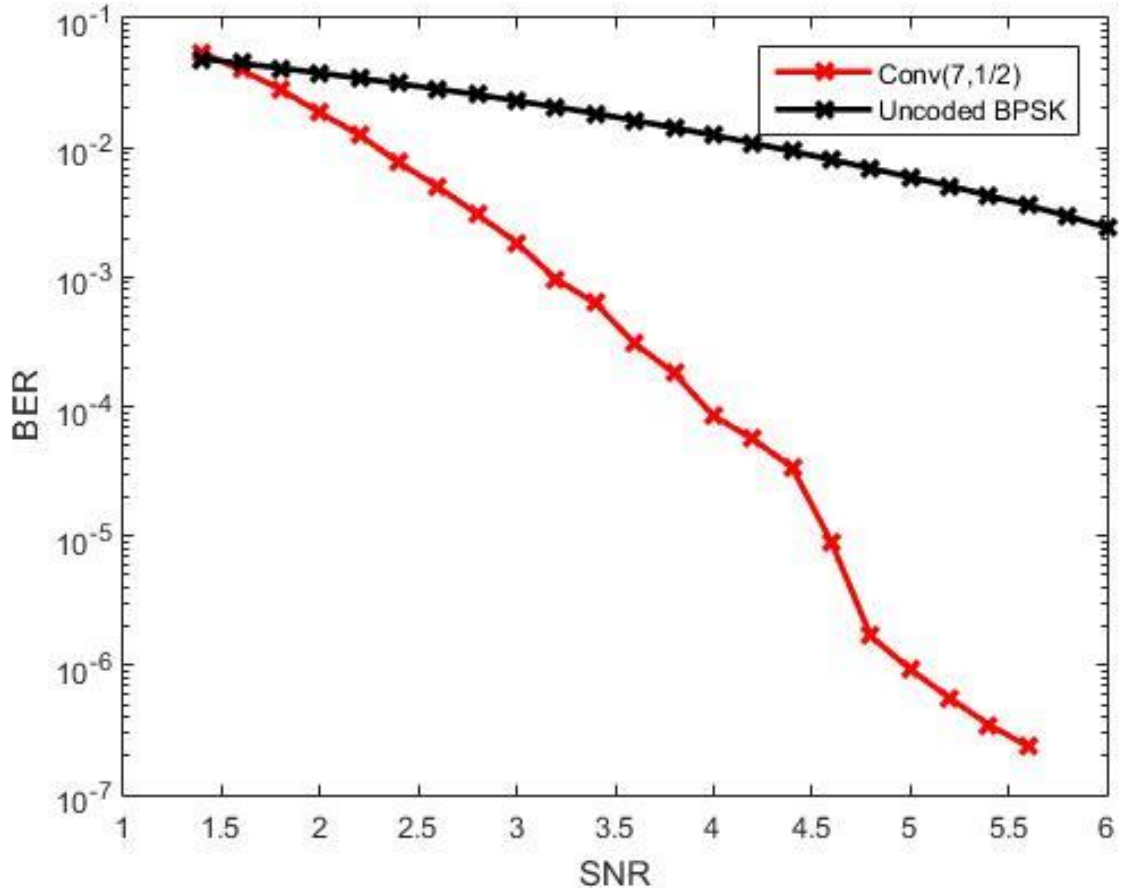

The increase in performance is huge as expected compared to an uncoded transmission. Even on a quite low SNR value of 1.5dB the convolutional code is able to produce a better performance than the uncoded channel. As the SNR increases the performance of the transmission using the convolutional code is getting much better than the uncoded channel having a gain of about 5 dB to achieve a BER of $2 \cdot 10^{-7}$. The convolutional code produces an impressive gain for each increase of 0.2dB which lasts from the very low SNR to high values of it. Having used the uncoded transmission we would need twice the power used to accomplish the same results. So the convolutional code is an outstanding choice for our purpose with relatively low complexity and latency.

### 5.3 Simulation of the Reed Solomon code

The characteristics of the RS code that was used are:

- The message part is 223 symbols

- 2t=32 symbols for parity check, so t=16

- So 255 symbols per codeword

The field generator polynomial is

$$F(x) = x^8 + x^7 + x^2 + x + 1$$

(5.1)

over GF(2). The code generator polynomial is:

$$g(x) = \prod_{j=128-t}^{127+t} (x - a^{11j}) = \prod_{j=112}^{143} (x - a^{11j})$$

This is over $(2^8)$, where $F(a) = 0$.

(5.2)

So $a_g = a^{11}$, FR=112 and the biggest number/symbol is 255.

**Figure 5-3: Performance of the RS(255,223) code**

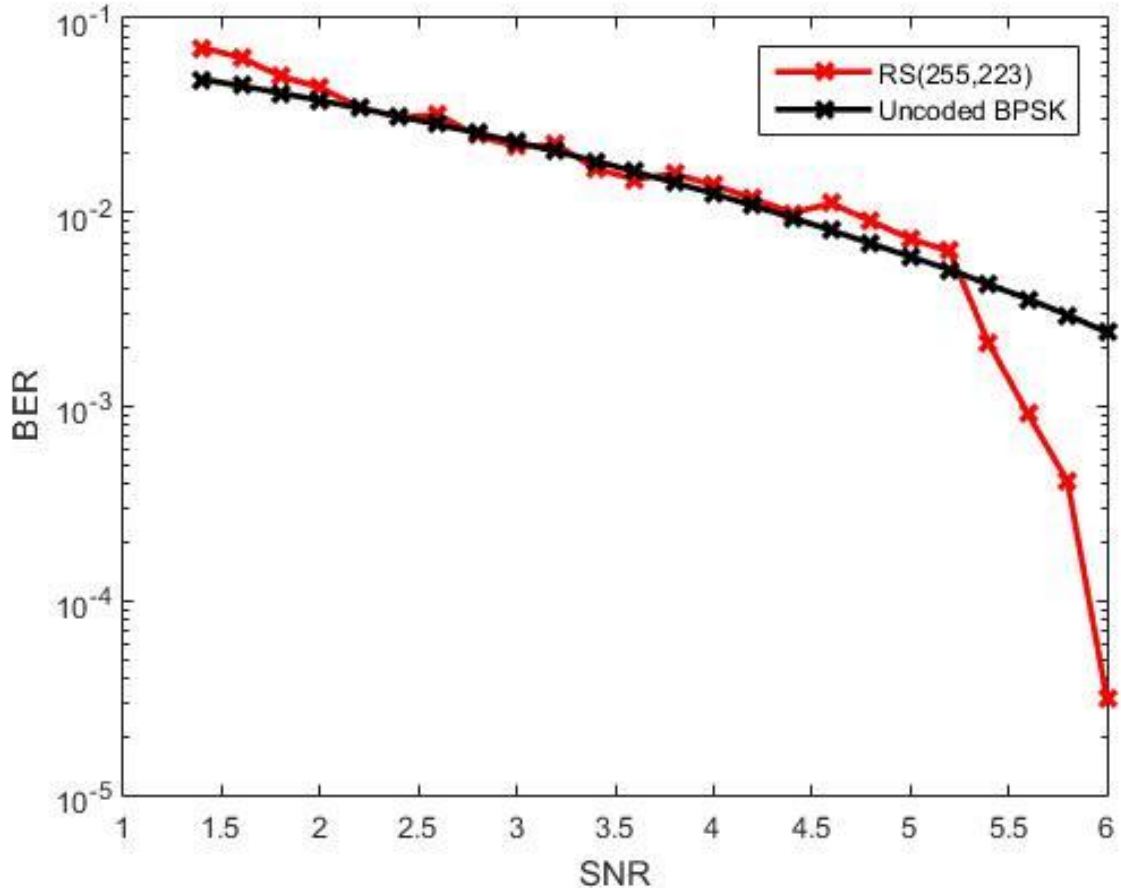

Figure 5-3 illustrates the performance of the proposed Reed Solomon code compared to the uncoded BPSK transmission over an AWGN channel. The figure shows that only on a SNR higher than 5dB the encoding process produces a considerable effect. The bound of the proposed code is t=16, so it cannot correct an error if more than 16 have occurred on the same codeword. Even with a quite big parity check number of symbols, the errors become sparse enough to be corrected only on a high SNR. But when the SNR reaches values as high as 5.2dB we notice a considerable decrease in the BER of the Reed-Solomon code which continues to decrease sharply to achieve an important gain. Specifically a value of $3 \cdot 10^{-5}$ BER for a SNR of 6dB is accomplished while the uncoded version has a BER of $3 \cdot 10^{-2}$ for the same SNR value, resulting in a $x10^3$ better BER for the same transmitting power.

## 5.4 Simulation of the concatenated code

One widespread method to build a strong code while maintaining manageable decoding complexity in space communications is to concatenate two codes, an 'outer code' and an 'inner code'. The proposed coding system consists of the Reed-Solomon outer code and the convolutional inner code (which is Viterbi decoded). Typically, the inner convolutional code corrects enough errors so that a high-code-rate outer code can reduce the error probability to the desired level. The convolutional decoder usually produce errors in short bursts. This makes the Reed Solomon code ideal to correct, since short burst errors will mostly occur in the same symbol which will account as just one error of the 16 that are correctable. The proposed RS(255,223) code's symbols consist of 8 bits each, so up to 8 consecutive errors produced by the Viterbi decoder will be clustered in one Reed Solomon symbol.

### 5.4.1 System without an interleaver

Figure 5-4 depicts the concatenated system used.

**Figure 5-4: The structure of the concatenated system**

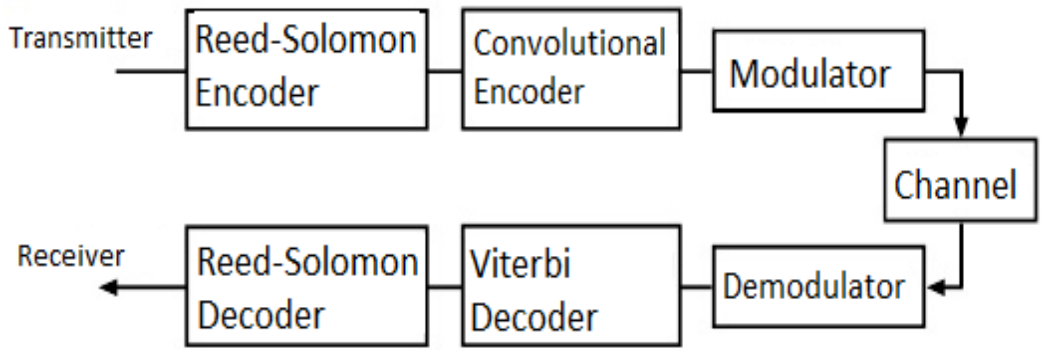

**Figure 5-5: Performance of the proposed concatenated system**

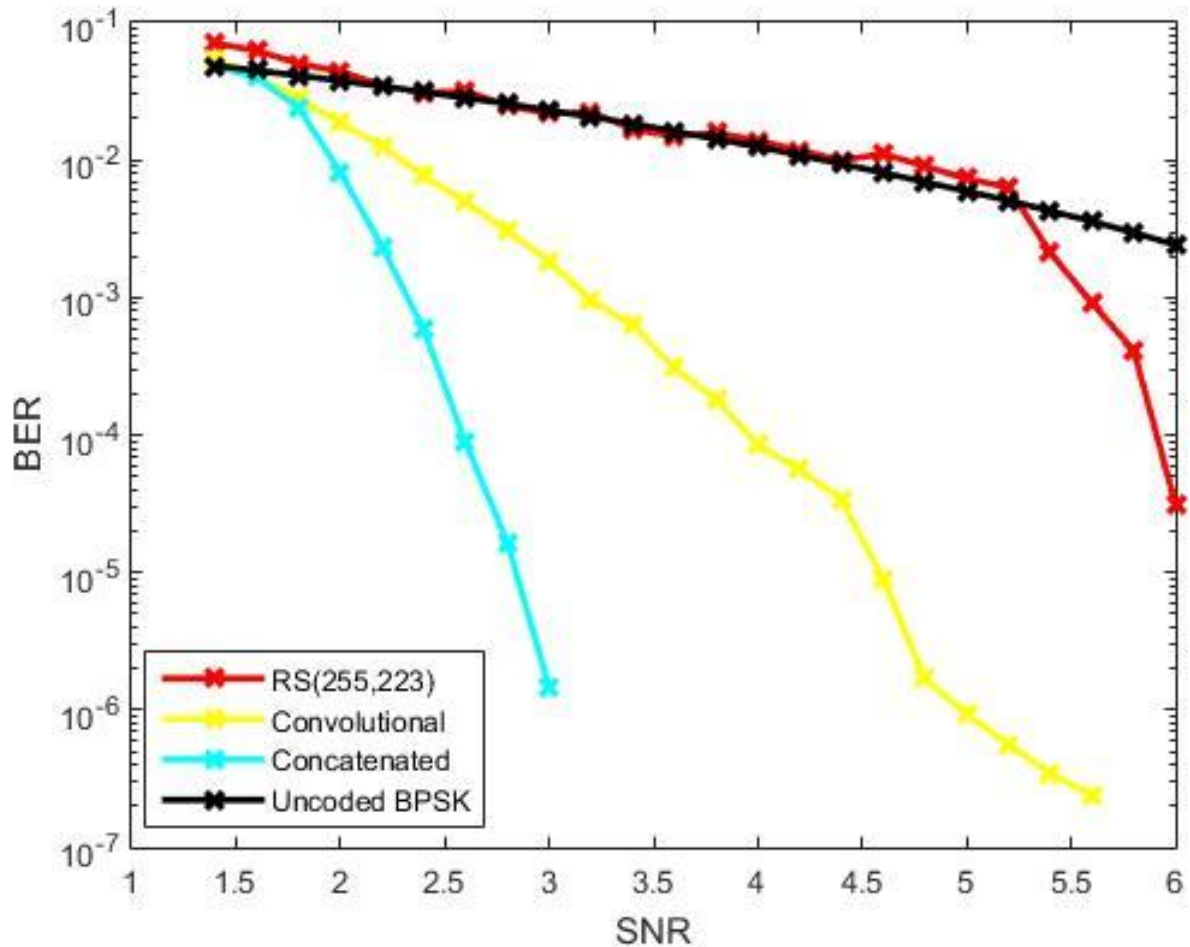

Figure 5-5 illustrates the performance of the proposed concatenated system coding for an AWGN channel and BPSK modulation. We can easily notice the great performance gain received from the concatenation of the two coding methods. The performance of the concatenated coder shows a coding gain of more than 1.5 dB from the best until now convolutional code. The performance gain of the concatenated system is outstanding compared to the uncoded transmission specifically it approaches a BER of $10^{-6}$ in a SNR of 3dB this performance is approached by the uncoded version with a SNR more than three times of that. Also the comparison between the concatenated an the Reed-Solomon code gives a coding gain of more than 3dB. So it is essential to use such a system provided that it produces a significant performance compared to any of the other versions of the same simulated transmission.

### 5.4.2 System with an interleaver

The concatenated system in fig.5-4 produces a significant increase in the performance of the transmission but the gain can be increased even more by inserting an interleaver after the Reed-Solomon coder. This will have as a result for the error bursts, which the Viterbi decoder produces, to be spread to different Reed-Solomon codewords decreasing the number of erroneous symbols in each individual codeword, thus improving the error correction efficiency.

The function of the interleaver is very easy. It just buffers the input symbols row by row, but outputs them column by column. So an interleaver of length 2 will output firstly the $1^{st}$ symbol of the $1^{st}$ codeword and secondly the $1^{st}$ symbol of the $2^{nd}$ codeword, then the $2^{nd}$ symbol of the $1^{st}$ codeword and so on as shown in fig5-6.

**Figure 5-6:Interleaver's function**

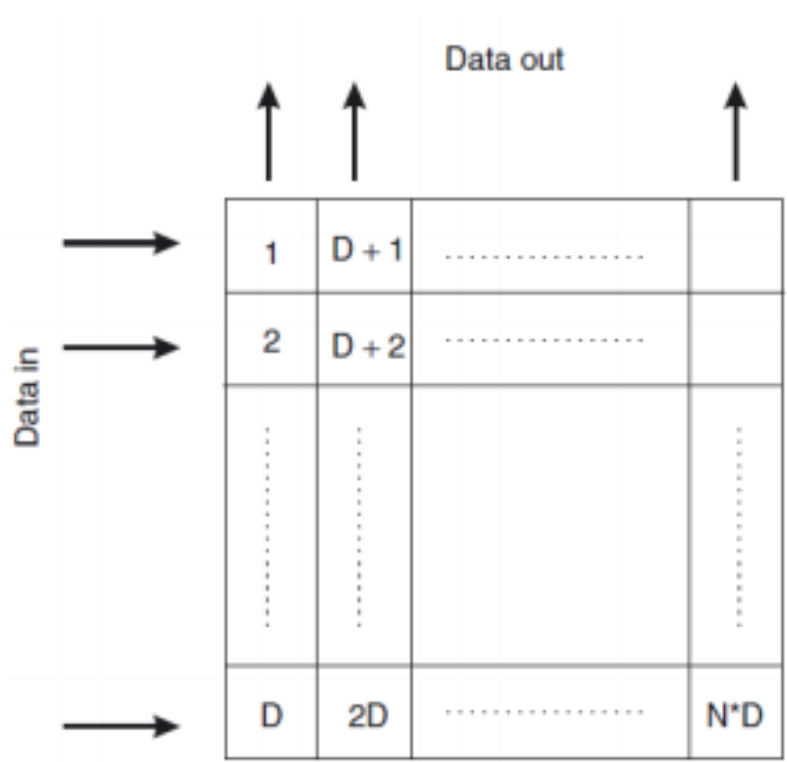

This functions spreads the symbols to different timeslots so that when an error burst occurs the damage for each codeword will be small and recoverable. However, this technique requires to delay transmission and reception since we need D=interleaving depth codewords to be able to decode each message.

 Figure 5-7 shows the new system with the interleaver included.

**Figure 5-7: The system including the interleaver**

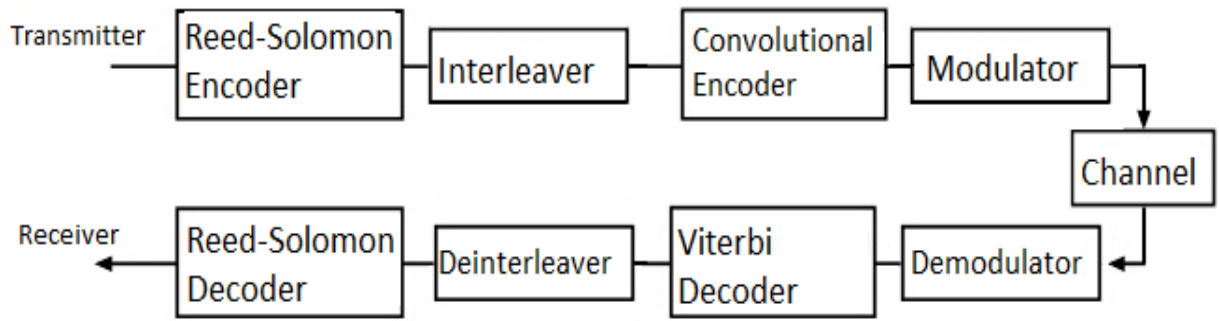

The effect of the inclusion of the interleaver can be noticed in fig.5-8 which shows the performance for the concatenated system using different depths of interleaving. The function of the interleaver is so significant that it can produce coding gain of about 0.5dB. As expected the higher the interleaving depth the better the performance is. We can say that D=5 is an optimal value since it approaches the ideal performance, without much delay as higher depths require to wait more for the whole D codewords to arrive. To perform this simulation the same messages were sent but they were "cut" and sent in different portions in order to achieve the recommended interleaving. We can see the result of this technique and it is easy to understand that the more we spread the parts of the codewords the more rarely these parts will be erroneous since errors occur in small bursts and it will unlikely to hit many symbols of the same codeword.

**Figure 5-8: Performance of the concatenated system for different interleaving depths**

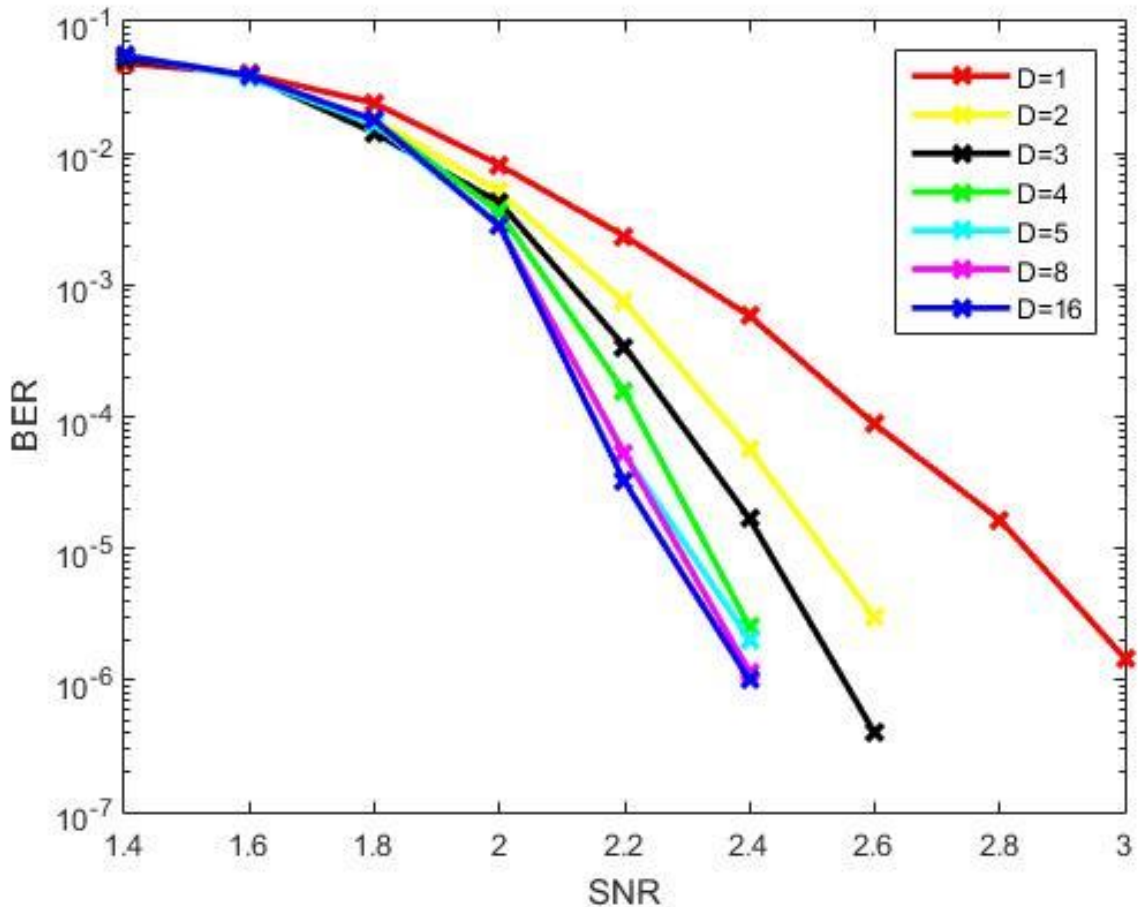

### 5.4.3: Overall coding methods comparison

Figure 5-9 illustrates a comparison among all the different coding techniques that were used to have a complete view of what performance each of them had and how much it was increased after applying the full coding system illustrated in figure 5-7. An incredible gain of about 4dB is noticed compared to the convolutional code, the best performer after the concatenated system. Comparing the final system that was implanted to the uncoded transmission we can see that a gain of about 4x in dB is produced. In conclusion, the final proposed system proves to be a very efficient one and able to achieve a reliable communication in very bad conditions, so it can be put in use for the many and severe difficulties met in space communications.

**Figure 5-9: Performance comparison of different coding methods**

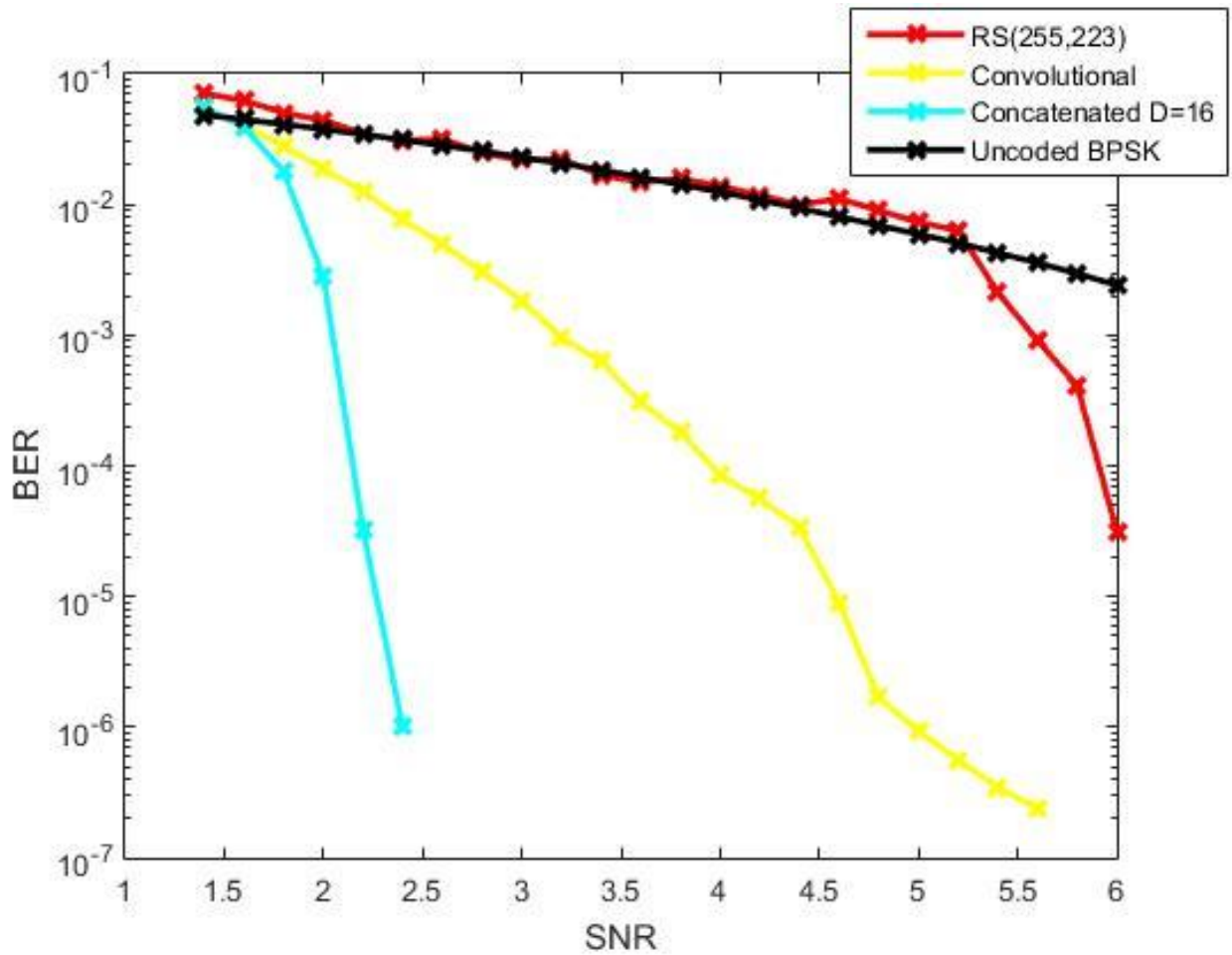

## 6. CONCLUSION

In this paper the concatenated CCSDS coding system was studied. The whole encoding and decoding process was presented after first explaining the required algebra background. First the Reed-Solomon encoder is examined and developed in Matlab as it was easier to understand and then the decoder of this code is presented with the different approaches that exist to complete each step of the decoding process. Afterwards, the convolutional code's encoder and decoder is demonstrated and implemented for our specific implementation. After successfully implementing each code separately their concatenation is easily constructed by having the Reed-Solomon as an outer code and the convolutional as an inner code.

Through this process the complexity of each step of the transmission is realized. We can see that there are different ways to compute the required elements but the steps that it consists of are constant and come in the same order to successfully transmit a message.
In addition to that, we demonstrate the performance gain that this approach produces compared to using each coding method separately and the huge gain that the inclusion of an interleaver can yield.